\documentclass[journal]{IEEEtran}

\IEEEoverridecommandlockouts
\usepackage{amsmath,amssymb,amsfonts,accents,mathrsfs}
\usepackage{textcomp}
\def\BibTeX{{\rm B\kern-.05em{\sc i\kern-.025em b}\kern-.08em
    T\kern-.1667em\lower.7ex\hbox{E}\kern-.125emX}}

\usepackage{ulem}
\usepackage{float}
\usepackage{bm}

\usepackage{graphicx}
\usepackage{textcomp}
\usepackage[dvipsnames]{xcolor}

\definecolor{amethyst}{rgb}{0.6, 0.4, 0.8}

\usepackage{tikz}
\usetikzlibrary{positioning}
\usepackage{textcomp}
\usetikzlibrary{shapes,arrows,backgrounds}
\usetikzlibrary{datavisualization}
\usetikzlibrary{patterns}
\usepackage{verbatim}

\usepackage{pgfplots}
\pgfplotsset{compat=newest}
\pgfplotsset{plot coordinates/math parser=false}
\newlength\figureheight
\newlength\figurewidth

\definecolor{x11_gray}{rgb}{0.85, 0.85, 0.85}
\definecolor{darkgreen}{rgb}{0.0, 0.5, 0.0}
\definecolor{amethyst}{rgb}{0.6, 0.4, 0.8}

\usepackage[labelsep=quad,indention=10pt]{subfig}
\usepackage{algorithm}
\usepackage{algorithmic}

\usepackage{array}

\usepackage{tabularx}

\usepackage{url}
\usepackage{mathtools}
\usepackage{cuted}

\renewcommand{\t}{\left[t\right]}
\newcounter{MYtempeqncnt}
\usepackage{nicefrac}

\begin{document}
\title{Iterative List Detection and Decoding for mMTC}%Massive Machine-Type Communications}

% Authors
% -------------------------------------------------------------------------
\author{R. ~B.~Di~Renna,~\IEEEmembership{Student Member,~IEEE,}
        and~R.~C.~de~Lamare,~\IEEEmembership{Senior Member,~IEEE}% <-this % stops a space
\thanks{The authors are with the Centre for Telecommunications Studies (CETUC),
Pontifical Catholic University of Rio de Janeiro (PUC-Rio), Rio de Janeiro 22453-900,
Brazil (e-mail: robertobrauer@cetuc.puc-rio.br; delamare@cetuc.puc-rio.br).
This work was supported in part by the National Council for Scientific and Technological Development (CNPq) and in part by FAPERJ.}% <-this % stops a space
\thanks{}}

%\markboth{IEEE TRANSACTIONS ON COMMUNICATIONS,~Vol.~XX, No.~X, Month~2020}%
%{Di Renna \MakeLowercase{\textit{et al.}}: Adaptive and Iterative Activity-Aware Variable Group-List Decision Feedback Detection for mMTC}

\maketitle

%%%%%%%%%%%%%%%%%%%%%%%%%%%%%%%%%%%%%%%%%%%%%%%%%%%%%%%%%%%%%%%%%%%%%%%%%%%%%%%%%
% ABSTRACT
%%%%%%%%%%%%%%%%%%%%%%%%%%%%%%%%%%%%%%%%%%%%%%%%%%%%%%%%%%%%%%%%%%%%%%%%%%%%%%%%%
\begin{abstract}
The main challenge of massive machine-type communications (mMTC) is the joint activity and signal detection of devices. The mMTC scenario with many devices transmitting data intermittently at low data rates and via very short packets enables its modelling as a sparse signal processing problem. In this work, we consider a grant-free system and propose a detection and decoding scheme that jointly detects activity and signals of devices. The proposed scheme consists of a list detection technique, an $l_0$-norm regularized activity-aware recursive least-squares algorithm, and an iterative detection and decoding (IDD) approach that exploits the device activity probability. In particular, the proposed list detection technique uses two candidate-list schemes to enhance the detection performance. We also incorporate the proposed list detection technique into an IDD scheme based on low-density parity-check codes. We derive uplink sum-rate expressions that take into account metadata collisions, interference and a variable activity probability for each user. A computational complexity analysis shows that the proposed list detector does not require a significant additional complexity over existing detectors, whereas a diversity analysis discusses its diversity order. Simulations show that the proposed scheme obtains a performance superior to existing suboptimal detectors and close to the oracle LMMSE detector.
\end{abstract}

%%%%%%%%%%%%%%%%%%%%%%%%%%%%%%%%%%%%%%%%%%%%%%%%%%%%%%%%%%%%%%%%%%%%%%%%%%%%%%%%%
% KEYWORDS

\IEEEpeerreviewmaketitle

%%%%%%%%%%%%%%%%%%%%%%%%%%%%%%%%%%%%%%%%%%%%%%%%%%%%%%%%%%%%%%%%%%%%%%%%%%%%%%%%%
% INTRODUCTION
%%%%%%%%%%%%%%%%%%%%%%%%%%%%%%%%%%%%%%%%%%%%%%%%%%%%%%%%%%%%%%%%%%%%%%%%%%%%%%%%%
\section{Introduction}
\label{sec:int}
% Contextualização
\IEEEPARstart{M}{assive} machine-type communications (mMTC) has been considered a technology with great potential in future networks. This potential can be widespread across different industries, including healthcare, logistics, manufacturing, process automation, energy, and utilities. Different from the conventional human type communications, mMTC for IoT have unique service features, as transmissions between two MTC devices, low data rates, very short packets and high requirements of energy efficiency and security~\cite{Popovski2018,Salam2019}.

In this context, grant-free access is a promising technique to meet the specifications of mMTC. Dividing the short packages in preamble (metadata) and payload (data), the central aggregation node can detect the active devices and estimate their channels in just one transmission. Given the multiple applications for mMTC, each type of device has its own activity behaviour. For example, monitoring devices of high-risk patients in a hospital request access to the network more frequently than gas sensors from a smart home. Thus, it is natural to assume that each device has a different probability of being active ($\rho_n$).

As mMTC is a very dense scenario and due to the coherence interval of the channel, even with a small number of active devices (10\% of the cell) the reuse of metadata sequences must occur. Naturally, metadata allocation has to adapt with the traffic activity pattern, designating the resources in a random manner~\cite{ECarvalho2017}. Despite many works~\cite{Hu2016,HYin2016,Adhikary2013,Bjorson2018} that consider the reuse only in the neighbouring cells, the assumption of intra-cell metadata contamination should be applied~\cite{Osseiran2014}. This assumption is directly related to the uplink capacity analysis. For this scenario, the sum-rate expressions besides take metadata contamination into account, must consider the different activity probability of each device.

As envisaged, this kind of network supports a massive number of devices with sporadic activity, this scenario can be interpreted as a sparse signal processing problem~\cite{Liu2018}. Exploiting the sparsity of the system, detection algorithms should be reformulated to the mMTC scenario. A common approach is to apply a regularization parameter into the cost function, as in~\cite{Zhu2011} wherein Zhu and Giannakis proposed the Sparse Maximum a Posteriori Probability (S-MAP) detection which performs a MAP detection of the new sparse problem, considering a zero-augmented finite alphabet. A variation of the well-known sphere decoder has been proposed in~\cite{Knoop2014} and named as K-Best. In~\cite{Knoop2013}, a sparsity constraint has been incorporated in the successive interference cancellation (SA-SIC). In order to avoid matrix inversions, in~\cite{Ahn2018} a version of SA-SIC with sorted QR decomposition and ordered detection based on the activity probability of devices has been reported. Despite its large computational complexity, a solution belonging to the class of Bayesian interference algorithms has been described in~\cite{Zhang2018}, which has the advantage of performing the detection without knowing the activity factor $\rho_n$.

% Descrição da proposta
In the approach proposed in this paper, when a device has to transmit data, it splits the codeword in multiple frames and transmit them in multiple transmission slots. During the same coherence time, the channel is constant and it is estimated from uplink metadata every time the device transmits. In each time slot, each active device selects randomly a metadata sequence from a predetermined codebook and sends the rest of the codeword. As the number of orthogonal metadata sequences is lower than the devices, the system is suitable to frame collisions~\cite{ECarvalho2017,Liu2018}.

In this work, inspired by the joint activity and data detection problem in sparse scenarios, we introduce a detection scheme named activity-aware variable group-list decision feedback (AA-VGL-DF). The proposed AA-VGL-DF scheme consists of a list detector, an $l_0$-norm regularized recursive least-squares (RLS) algorithm that exploits the sparsity of the system to adjust the receive filters, and an iterative detection and decoding (IDD) scheme. Inspired by our previous list-detection work, the AA-MF-SIC~\cite{DiRenna2019}, we employ two list detection techniques based on the constellation points, thus increasing the accuracy of each symbol detection. In order to reduce the computational complexity, we consider a variable size of each list of constellation points candidates based on the SINR. An IDD scheme based on low-density parity-check (LDPC) codes, which incorporates the $l_0$-norm regularized RLS algorithm, the list-detector and takes advantage of the activity probability of each device is also devised for signal detection in mMTC. We then derive uplink achievable sum-rate expressions that take into account metadata collisions, interference and a variable activity probability for each user.  An analysis of the computational complexity shows that the AA-VGL-DF detector does not require a significant additional complexity over existing techniques, whereas a diversity analysis discusses the diversity order achieved by the AA-VGL-DF detector. Simulations show that our AA-VGL-DF scheme successfully mitigates the error propagation and approaches the oracle linear minimum mean-squared error (LMMSE) detector performance with a competitive complexity.

% Principais contribuições do trabalho
The main contributions of this work are:
\begin{enumerate}
    \item The AA-VGL-DF detection scheme with an $l_0$-norm regularized RLS algorithm with two list-based strategies, which detects active devices and their symbols;
    \item An IDD scheme based on LDPC codes modified for mMTC that incorporates the AA-VGL-DF detector;
    \item A diversity order analysis along with a complexity analysis of AA-VGL-DF and existing approaches based on required floating-point operations (FLOPs);
    \item A derivation of a closed-form expression for the achievable spectral efficiency which takes metadata collisions, interference and the activity probability of each user;
    \item A comparative study with simulation results of the AA-VGL-DF and existing techniques.
\end{enumerate}

% organização do paper
The organization of this paper is as follows: Section~\ref{sec:syst_model} briefly describes the random access and the channel models. Section~\ref{sec:adapt_imp} presents the AA-VGL-DF detector, the $l_0$-norm regularization and the variable group-list constraint. Section~\ref{sec:prop_it_det_dec} introduces the IDD and details the modifications that are suitable for mMTC. Analyses of computational complexity,  diversity order and the achiveable uplink sum-rate are developed in Section~\ref{sec:analysis}. Section~\ref{sec:sim} presents the setup for simulations and results while Section~\ref{sec:Conc} draws the conclusions.

\textit{Notations:} Matrices and vectors are denoted by boldfaced capital letters  and lower-case letters, respectively. The space of complex (real) $N$-dimensional vectors is denoted by $\mathbb{C}^N\left(\mathbb{R}^N\right)$. The $i$-th column of a matrix $\mathbf{A} \in \mathbb{C}^{M\times N}$ is denoted by $\mathbf{a}_i \in \mathbb{C}^M$. The superscripts $\left(\cdot\right)^T$ and $\left(\cdot\right)^H$ stand for the transpose and conjugate transpose, respectively. For a given vector $\mathbf{x} \in \mathbb{C}^N, ||\mathbf{x}||$ denotes its Euclidean norm. $\mathbb{E\left[\cdot\right]}$ stands for expected value, $\mathbf{I}$ is the identity matrix and $\text{diag}\left[\cdot\right]$ is to reshape a vector in the main diagonal of a matrix.

%%%%%%%%%%%%%%%%%%%%%%%%%%%%%%%%%%%%%%%%%%%%%%%%%%%%%%%%%%%%%%%%%%%%%%%%%%%%%%%%%
% SYSTEM MODEL
%%%%%%%%%%%%%%%%%%%%%%%%%%%%%%%%%%%%%%%%%%%%%%%%%%%%%%%%%%%%%%%%%%%%%%%%%%%%%%%%%
\section{System Model} \label{sec:syst_model}
This section presents the sparse signal model and the system setup considered for the mMTC scenario. The sparse signal model takes into account considerations available in a 3GPP specification and the system setup details the received signal in a given coherence time.

%%%%%%%%%%%%%%%%%%  Random access model
\subsection{Sparse signal model}

As networks should support many different applications with distinct requirements, naturally each device has its own activity behaviour. In order to model this scenario, we consider the beta-binomial distribution~\cite{Skellam1948} to model the problem as the probability of being active $\rho_n$ is randomly drawn from a beta distribution, proposed as a traffic model by the 3GPP~\cite{3GPP_TR37868}. Considering $N$ devices with a single antenna accessing a base station (BS), the probability mass function is given by
\begin{equation}\label{eq:pdf}
    p\left(K\right) = \begin{pmatrix}N \\ K\end{pmatrix} \rho_n^K \left(1-\rho_n \right)^{N-K}
\end{equation}

\noindent where $\rho_n$ is a random variable with a beta distribution that represents the probability of being active of the $n$-th device and $K$ is the number of active devices out of $N$ at the same transmission slot.
The sparsity of the scenario is modified as soon as each random variable $\rho_n$ with beta distribution is modelled. Thus, each device has its own activity probability.

Hence, the probability of having K active devices within a total of N at the same transmission slot is given by

\begin{align}\nonumber
\hspace{-5pt}   p\left(K\,|\, N,\alpha,\beta\right) =& \frac{\Gamma\left(N+1\right)}{\Gamma\left(K+1\right)\Gamma\left(N-K+1\right)}\\
    & \frac{\Gamma\left(N+\alpha\right)\Gamma\left(N-K+\beta\right)}{\Gamma\left(N+\alpha+\beta\right)} \frac{\Gamma\left(\alpha+\beta\right)}{\Gamma\left(\alpha\right)\Gamma\left(\beta\right)}
\end{align}

\noindent where $\Gamma\left(\cdot\right)$ is the gamma function, $\alpha$ and $\beta$ are real positive parameters that appear as exponents of the random variable $\rho_n$ and control the shape of the distribution. The average number of active devices is $\frac{N \alpha}{\alpha + \beta}$ and its variance is $\frac{N\alpha \beta \left(\alpha + \beta + N\right)}{{\left(\alpha + \beta\right)}^{2}\left(\alpha + \beta +1\right)}$.

\begin{figure}[t]
    \centering
    \includegraphics[scale=0.9]{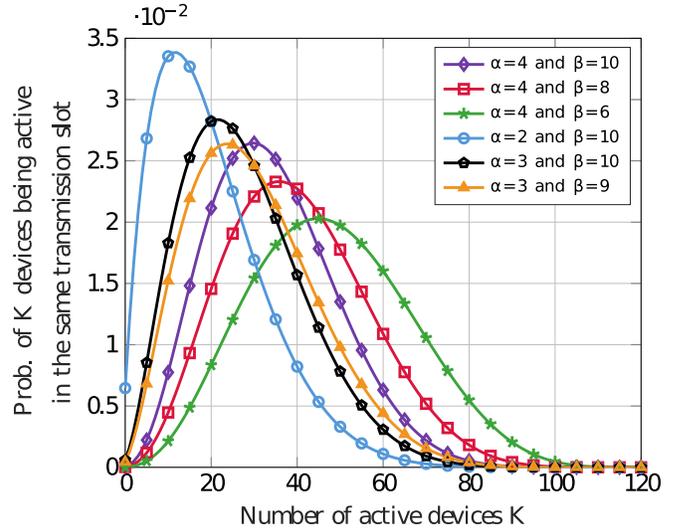}
    \caption{\footnotesize Model of probability of having $K$ active devices within a total of $N=120$. All devices become independently active with a probability determined by the random variable $\rho_n \in \left[0,0.1,\dots,1\right]$ with a beta distribution.}
    \label{fig:prob_act_devices}
\end{figure}

In most mMTC applications the devices have low probability of being active. Fig.~\ref{fig:prob_act_devices} shows the probability of a specific number of devices to be active at the same time for different values of $\alpha$ and $\beta$ in a scenario of $N=120$.

\subsection{System setup}
\begin{figure}[t]
    \begin{center}
        \includegraphics[scale=0.35]{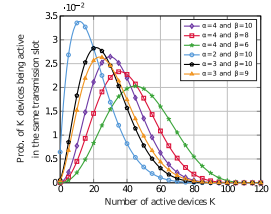}
        \caption{\footnotesize mMTC single-cell system model. When active, each device transmits $\tau_\phi$ and $\tau_x$ symbols of metadata and data, respectively, in the coherence time.}
        \label{fig:mtc_sys}
    \end{center}
\end{figure}

The massive uplink connectivity scenario illustrated in Fig.~\ref{fig:mtc_sys} is considered, where $N$ devices with a single antenna access a single base station (BS), equipped with $M$ antennas. When a device has data to transmit, it splits the codeword in multiple frames and transmit them in multiple transmission slots. In each time slot, each active device selects randomly a metadata sequence from a predetermined codebook and sends the rest of the codeword. Since in practice the BS would have a list of devices that are associated with it, and their unique identifiers, we assume that the metadata sequences are known at the BS. Since these unique identifiers are known to the BS, the metadata sequence is also known at the BS. Given the sporadic activity of devices, they will communicate to the BS only when it is needed, so not all of them will be active during the same coherence time.

As the frame size of mMTC is typically very small (between 10 and 100 bytes)~\cite{Salam2019}, it is possible to assume the devices are synchronized in time. That is, the devices are turned on or turned off in the same transmission slot, as represented in Fig.~\ref{fig:mtc_sys}. The duration of a transmission slot ($\tau = \tau_\phi+\tau_x$) is smaller than the coherence time and coherence bandwidth of the channel. The time index $t$ indicates each transmitted vector in the same transmission slot. As we considered a grant-free random access model, each frame has metadata and data. Thus, the time index indicates how each frame is divided, as in Fig.~\ref{fig:mtc_sys}.

The received signal $\mathbf{y}\t$ in a given coherence time is organized in a $M \times 1$ vector that contains the transmitted metadata ($\boldsymbol{\phi}\t$) or the data ($\mathbf{x}\t$), as

\begin{equation}\label{eq:f_sumY1}
    \mathbf{y}\t =\left\{
    \begin{array}{ll}
         \mathbf{H} \, \sqrt{\tau_\phi} \,\mathbf{B} \,\boldsymbol{\phi}\t + \mathbf{v}\t,& \text{if}\hspace{10pt} 1 \leq t \leq \tau_\phi  \\
         \mathbf{H} \, \sqrt{\tau_x} \, \mathbf{B} \hspace{3pt}\mathbf{x}\hspace{1pt}\t + \mathbf{v}\t ,& \text{if}\hspace{10pt}  \tau_\phi  < t \leq \tau
    \end{array}\right.
\end{equation}

\noindent where $\mathbf{H}$ is the $M\times N$ channel matrix, $\mathbf{B}$ is the $N\times N$ transmission power matrix, $\mathbf{v}$ is the $M\times 1$ noise vector, while $\tau_\phi$ and $\tau_x$ are the number of metadata and data symbols, respectively. For each time instant $t$, the metadata and data are represented by the $N \times 1$ vectors
\hspace{-5pt}
\begin{eqnarray}
    \boldsymbol{\phi}\t &\hspace{2.2pt}= \hspace{8pt} \boldsymbol{\Delta}\,\boldsymbol{\varphi}\t &= \left[\delta_1\,\varphi_{1}\t,\dots,\delta_N\,\varphi_{N}\t\right]^T\hspace{2pt} \text{and} \\
    \mathbf{x}\t &= \hspace{8pt} \boldsymbol{\Delta}\,\mathbf{s}\t &= \left[\delta_1\,s_{1}\t,\dots,\delta_N\,s_{N}\t\right]^T,
\end{eqnarray}

\noindent where $\boldsymbol{\varphi}\t$ and $\mathbf{s}\t$ are $N \times 1$ vectors of symbols from a regular modulation scheme denoted by $\mathcal{A}$, as quadrature phase-shift keying (QPSK). The $N \times N$ diagonal matrix $\boldsymbol{\Delta}$ controls each device activity in the specific transmission slot, with Pr$\left(\delta_n = 1\right) = \rho_n$ and Pr$\left(\delta_n = 0\right) = 1-\rho_n$. Thus, each transmitted vector ($\boldsymbol{\phi}\t$ \text{or} $\mathbf{x}\t$) is composed by the augmented alphabet $\mathcal{A}_0$, where $\mathcal{A}_0 = \mathcal{A} \cup \left\{0\right\}$.

As in mMTC systems the transmission power $b$ of each device is different~\cite{Popovski2018},~\hspace{-4.25pt}\cite{Salam2019}, we gather the transmission power component in the $N \times N$ diagonal matrix $\mathbf{B} = \text{diag}\left(\left[\sqrt{b_1}, \sqrt{b_2}, \dots, \sqrt{b_N}\right]\right)$. The noise vector $\mathbf{v}$ is modelled as an independent zero-mean complex-Gaussian ${M \times 1}$ vector with variance $\sigma^2_v$.

In our work, we consider the block fading model, where a channel realization is constant across a transmission slot duration and changes independently from slot to slot. The $M\times N$ channel matrix $\mathbf{H}$ corresponds to the channel realizations between the BS and devices, modelled as

\begin{equation}
    \mathbf{H} = \mathbf{A}\, \boldsymbol{\mathcal{N}}^{1/2},
\end{equation}

\noindent where $\mathbf{H}$ gathers independent fast fading, geometric attenuation and log-normal shadow fading. $\mathbf{A}$ is the $M\times N$ matrix of fast fading coefficients circularly symmetric complex Gaussian distributed, with zero mean and unit variance. The $N \times N$ diagonal matrix $\bm{\mathcal{N}}$ models the path loss and shadowing experienced by each device and is modelled as $10\log_{10}\left(\chi\right)+\omega$, where $\chi$ is the signal-to-noise ratio (SNR) and $\omega$ is a Gaussian random variable with zero mean and variance $\sigma^2_\omega$~\cite{Rappaport2002}. Thus, each vector $\mathbf{h}_{n}$ can be written as

\begin{equation}\label{eq:channel}
    \mathbf{h}_{n} = \mathbf{a}_{n}\sqrt{\eta_n},\hspace{10pt} \forall n = 1, \dots , N.
\end{equation}

\noindent The $\eta_n$ coefficients are assumed to be known at the BS and changes very slowly, reaching a new value just in a new transmission slot. All signal model parameters are described in Table~\ref{tab:para_sys}.

Given the features of mMTC scenarios, the number of devices $N$ is larger than that of antennas $M$ at the base station, in a way that it consists of an underdetermined system. However, the transmitted symbols can be detected as their vectors have a sparse structure as the rows corresponding to the inactive users are zero. So, the activity detection problem is reduced to finding the non-zero rows of $\boldsymbol{\phi}\t$.

The motivation of this work is to propose an efficient detection technique for mMTC. Conventional detection techniques are not suitable to deal with the small coherence time and limitation of the orthogonal metadata sequences. Thus, we present an iterative and adaptive detection technique that exploits the sparsity of the system and, using the specifications provided for the mMTC scenario, jointly detects the activity and data of devices, outperforming existing approaches.

%%%%%%%%%%%%%%%%%%  Table with signal model parameters
\begin{table}[t]
    \centering
    \caption{Description of signal model parameters.}
    \label{tab:para_sys}
    \begin{tabular}{lp{6.5cm}} \\ \hline
        Parameter           & Description \\ \hline
        $M$                 & Number of base station antennas; \\
        $N$                 & Number of devices; \\
        $K$                 & Number of active devices; \\
        $\tau$              & Number of transmitted symbols per trans. slot, given by $\tau = \tau_x + \tau_{\phi}$, where $\tau_{x}$ represents the data and $\tau_{\phi}$ the metadata;\\
        $\rho_n$            & Random variable with a beta distribution that represents the probability of being active of the $n$-device; \\
        $\mathbf{y}\t$      & $M \times 1$ received symbol vector of the time instant $t$; \\
        $\boldsymbol{\phi}\t$   & $N \times 1$ metadata vector of the time instant $t$ composed by the augmented alphabet $\mathcal{A}_0$;\\
        $\mathbf{x}\t$  & $N \times 1$ data vector of the time instant $t$ composed by the augmented alphabet $\mathcal{A}_0$;\\
        $\boldsymbol{\Delta}$ & $N \times N$ diagonal matrix that controls each device activity in the specific transmission slot; \\
        $\mathbf{B}$      & $N \times N$ diagonal matrix that gathers the transmission power of each device;  \\
        $\mathbf{v}\t$      & noise component, modelled as a independent zero-mean complex-Gaussian $M \times 1$ vector with variance $\sigma^2_v$;\\
        $\mathbf{H}$            & $M \times N$ channel matrix, where $\mathbf{H} =\mathbf{A}\, \boldsymbol{\mathcal{N}}^{1/2}$; \\
        $\mathbf{A}$ &$M \times N$ matrix of fast fading coefficients; \\
        $\boldsymbol{\mathcal{N}}$ &$N \times N$ diagonal matrix that gathers the path loss and shadowing experienced by each device; \\
        \hline
    \end{tabular}
\end{table}

%%%%%%%%%%%%%%%%%%%%%%%%%%%%%%%%%%%%%%%%%%%%%%%%%%%%%%%%%%%%%%%%%%%%%%%%%%%%%%%%%
% PROPOSED DETECTION
%%%%%%%%%%%%%%%%%%%%%%%%%%%%%%%%%%%%%%%%%%%%%%%%%%%%%%%%%%%%%%%%%%%%%%%%%%%%%%%%%
\section{Variable Group-List Decision Feedback Detection} \label{sec:adapt_imp}

\begin{figure*}[t]
    \centering
    \includegraphics[scale=0.27]{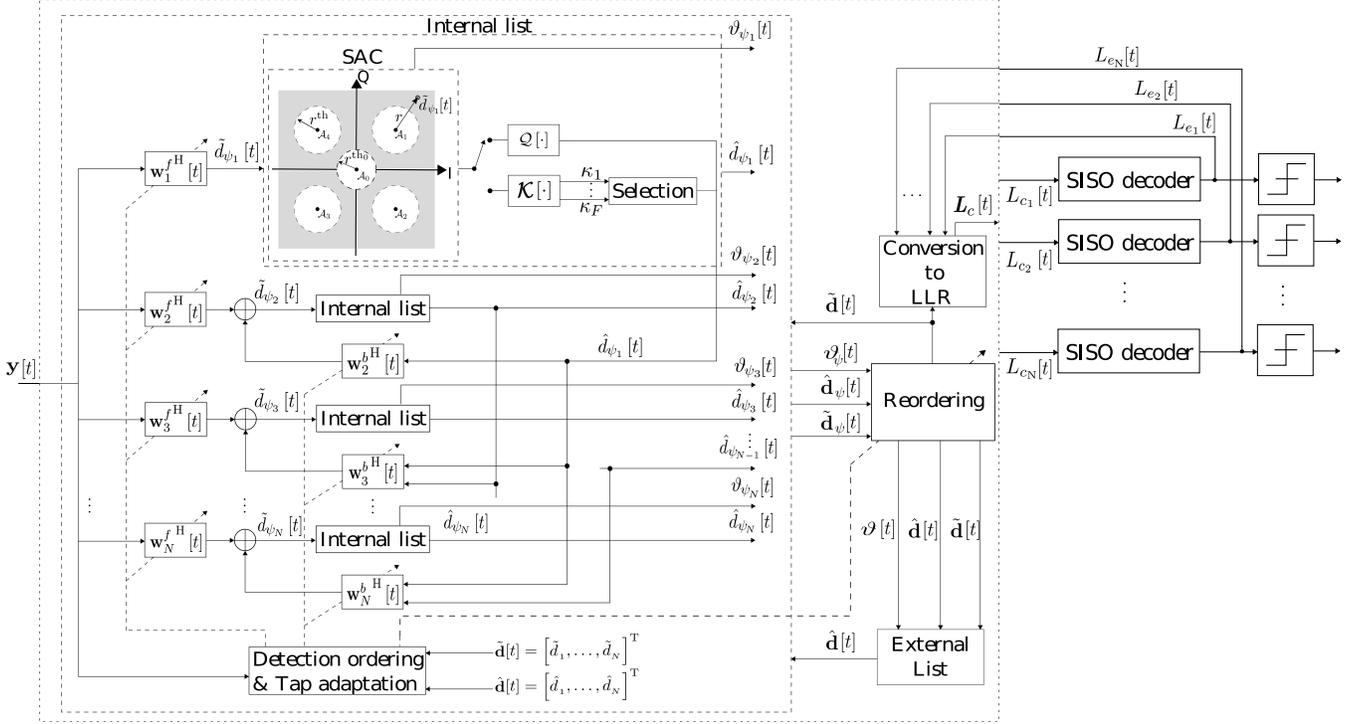}
    \caption{\footnotesize Detailed structure of the AA-VGL-DF detector and the IDD scheme. To simplify notations, just one received vector is considered in the base station.}
    \label{fig:bd_aavgldf}
\end{figure*}

This section details the proposed AA-VGL-DF detection scheme. Unlike
our previous work~\cite{DiRennaWCL2019}, AA-VGL-DF consists of an
adaptive receive filter adjusted by $l_0$-norm regularized RLS with
decision feedback and two list detection techniques to reduce the
detection error propagation. The detection scheme is illustrated in
a block diagram in Fig.~\ref{fig:bd_aavgldf}. Alternative detection
techniques
\cite{mmimo,wence,deLamare2003,itic,deLamare2008,cai2009,jidf,jiomimo,Li2011,
wlmwf,dfcc,deLamare2013,did,rrmser,bfidd,1bitidd,aaidd,aalidd} could
be examined and compared against AA-VGL-DF.

As the transmission slots are separated by metadata and data,
AA-VGL-DF has two modes of operation, training mode ($1 \leq t \leq
\tau_\phi$) and decision-directed mode ($\tau_\phi < t \leq \tau$).
Although the same scheme is used for both cases, in the training
mode the focus is to use the metadata to update the RLS algorithm,
while the decision-directed mode uses the filter to detect the
received data symbols. AA-VGL-DF detects each symbol at a time, per
layer. The detection order is updated at each new layer, using the
least squares estimation (LSE) criterion. The adaptive receive
filter can be decomposed into feedforward and feedback filters. The
feedforward one is updated at every new received vector by the
$l_0$-norm regularized RLS algorithm. The feedback filter is a
component that is concatenated to the feedforward filter in order to
cancel the interference of the previously detected symbols.

In the first layer, the filter is composed by the feedforward part
and obtains a soft estimate of the symbol with the received vector
$\mathbf{y}\t$. Inspired by our previous work~\cite{DiRenna2019}, we
apply the shadow area constraints (SAC) criterion to evaluate the
reliability of the hard decision of the soft symbol estimate. If the
symbol is considered unreliable, a list of candidates drawn from the
constellation symbols is generated and the best candidate, chosen by
the maximum likelihood criterion, replaces the unreliable symbol.
After the first detection, the LSE defines which will be the next
layer, the received vector $\mathbf{y}\t$ is concatenated with the
previous detected symbol and the filter of the selected layer is
concatenated with the feedback part. This procedure is repeated
until the last layer. At this point, the scheme has three vectors,
one with all soft estimates ($\tilde{\mathbf{d}}_\psi\t$), the
second one with the detected symbols ($\hat{\mathbf{d}}_\psi\t$) and
the third one ($\boldsymbol{\vartheta}_\psi\t$) that keeps the
information about the reliability of each soft estimate. Those three
vectors are reordered to the original sequence and the second list
of candidates begins. The idea of this list is to perform a group
list verification of the most unreliable symbols, after the last
detection. With more reliable detected vectors, a more accurate
decision of the first filter to be used in the next detection will
be taken. In order to define which soft estimates will be rechecked
with the second list, we apply the SAC criterion again, but with a
larger radius. After the reordering process, if AA-VGL-DF is in the
decision-directed mode, the reordered soft estimation vector
($\tilde{\mathbf{d}}\t$) is converted to LLRs in order to be decoded
by the iterative scheme. Otherwise, the reordered vectors of soft
estimates and detected symbols ($\hat{\mathbf{d}}\t$), are used to
define the first symbol to be detected in the next received metadata
vector.

We will present the proposed AA-VGL-DF scheme in detail in the following subsections. We first detail the adaptive decision feedback structure, the receive filters and the received vector concatenation. Then, we describe the internal verification list scheme and the detection order update. Lastly, we present the external list scheme and the $l_0$-norm regularized RLS algorithm.

%%%%%%%%%%%%%%%%%%  Block diagram

%%%%%%%%%%%%%%%%%%%%%%%%%%%%%%%%%%%%%%%
\subsection{Adaptive Decision Feedback Structure}\label{subsec:adapt_imp}

The main idea is to use a feedforward filter to detect the
transmitted symbol and a feedback filter to cancel the interference.
For each operation mode, both filters are concatenated and written
as
    \begin{equation}\label{eq:filter_forw_fedd}
        \mathbf{w}_{\psi_n}\t =
        \left\{
        \begin{array}{cl}
            \mathbf{w}^{f}_{\psi_n}\t, & n=1; \\
            \left[{\mathbf{w}^{f}_{\psi_n}}^\text{T}\t,{\mathbf{w}^{b}_{\psi_n}}^\text{T}\t\right]^\text{T}, &  n=2,\dots, N.
        \end{array}\right.
    \end{equation}
\noindent  where $\mathbf{w}_{\psi_n}\t$ corresponds to both filters used for the detection of the symbol of the $n$-th device (or layer). Both filters update their weights and the detection order $\boldsymbol{\psi}$ at each new symbol detection. The received vector $\mathbf{y}\t$ is concatenated with the $N \times 1$ vector $\hat{\mathbf{d}}_{\psi_{n-1}}\t$ which contains the previously detected symbols as
%\vspace{-5pt}
    \begin{equation}\label{eq:all_y}
        \mathbf{y}_{\psi_n}\t =
        \left\{
        \begin{array}{cl}
            \mathbf{y}\t, & n=1; \\
            \left[{\mathbf{y}}^\text{T}\t,{\hat{\mathbf{d}}^\text{T}_{\psi_{n-1}}}\t\right]^\text{T}, & n=2,\dots, N
        \end{array}\right.
    \end{equation}

\noindent and each soft symbol estimate of the $n$-th device is given by

    \begin{equation}\label{eq:out_equa}
        \tilde{d}_{\psi_n}\t = \mathbf{w}_{\psi_n}^\text{H}\t\mathbf{y}_{\psi_n}\t.
    \end{equation}

Thus, the filter $\mathbf{w}_{\psi_n}\t$ and the received vector $\mathbf{y}_{\psi_n}\t$ increases in length at each detection. In the last detection, $\mathbf{w}_{\psi_N}\t$ and $\mathbf{y}_{\psi_N}\t$ are a $\left(M+N\right) \times 1$ vectors, as $\mathbf{w}_{N}\t = \left[w^{f}_{\psi_N,1}\t, \dots, w^{f}_{\psi_N,M}\t, w^{b}_{\psi_N,M+1}\t,\dots, w^{b}_{\psi_N,M+N}\t\right]^\text{T}$.

\subsubsection{Internal list}
In order to improve the detection performance, we include a SAC, first presented in our previous work~\cite{DiRenna2019}, to evaluate the reliability of the soft estimates. As shown in Fig.~\ref{fig:bd_aavgldf}, with the augmented alphabet of a QAM modulation scheme, SAC compares the distance between the soft estimate and all the possible constellation symbols with
    \begin{equation}\label{eq:detec_ord}
        r = \underset{i\,\in \, 1,\cdots,\left(|\mathcal{A}|+1\right)}{\textrm{arg min}}\hspace*{5pt} \|\mathcal{A}_{0_i}-\tilde{d}_{\psi_n}\t\|^2,
    \end{equation}
\noindent where $|\mathcal{A}|$ is the modulation order. If the soft estimate falls into the shadow area ($r > r^\text{th}$ or $r > r^{\text{th}_0}$), the estimate is considered unreliable and then $\tilde{d}_{\psi_n}\t$ proceeds to the list scheme. Otherwise, it is just quantized to the nearest symbol of the augmented constellation $\mathcal{A}_0$, as
    \begin{equation}\label{eq:out_equa2}
        \hat{d}_{\psi_n}\t = \mathcal{Q}\left[\mathbf{w}_{\psi_n}^\text{H}\t\mathbf{y}_{\psi_n}\t\right].
    \end{equation}
The $r^\text{th}$ and $r^{\text{th}_0}$ radius of each reliability region are defined by the probability of being active of each device~\cite{DiRenna2019} and the radius of the region around the zero (inactive device) is the complement of the radius of the regions around the constellation symbols, $r^{\text{th}_0} = 1 - r^\text{th}$.

The list scheme is a verification of a list of candidates drawn from constellation symbols to the actual detection. The list $\boldsymbol{\kappa} = \left[\kappa_1\, \cdots, \kappa_{\left(|\mathcal{A}_0|\right)}\right]$ is used to select the best candidate according to
\begin{equation}
        \boldsymbol{\kappa}_{\text{opt}} = \underset{i\,\in\, 1,\cdots,\left(|\mathcal{A}_0|\right)}{\textrm{arg min}}\hspace*{5pt} \|\mathbf{y}\t - \hat{\mathbf{h}}_{\psi_n}\, \kappa_i \|^2,
\end{equation}
\noindent where the vector $\hat{\mathbf{h}}_{\psi_n}$ contains the estimate of the channel between the device that performs symbol detection and the BS. As the channel estimation is not the focus of this work, we considered the well known linear MMSE (LMMSE) estimation. The estimate of each $\hat{\mathbf{h}}_{\psi_n}$ is detailed in Section~\ref{subsec:ul}.

The vector with minimum argument $\boldsymbol{\kappa}_\text{opt}$ indicates which candidate $\kappa$ will replace the quantized version of the unreliable soft symbol estimate $\tilde{d}_{\psi_n}\t$.

\subsubsection{Detection order update}

The metric chosen to update the detection order $\boldsymbol{\psi}$,
is the minimum LSE. At each symbol detection, we compute the
$l_0$-norm cost function $\mathcal{J}_{j}\t$ for the symbols that
were not detected yet. The set $S_j\t$ contains the index of the
remaining symbols to be detected and is updated with the output
$\psi_j$ of the cost function. Thus, the index of the chosen filter
is stored in the sequence of detection $\boldsymbol{\psi}$,
represented as
    %\vspace{-2.5pt}
    \begin{equation}\label{eq:detec_ord2}
        \psi_j = \underset{j\in S_j}{\textrm{arg min}}\hspace*{5pt} \mathcal{J}_{j}\t,
    \end{equation}

\noindent where the indicator $j$ in~(\ref{eq:detec_ord2}) belongs to the set $S_j=\{1,2,\dots,N\}-\{\psi_1,\psi_2,\dots,\psi_{n-1}\}$, which contains the index of not yet detected symbols. Hence, the output index of the cost function is removed by the set and this symbol is chosen to be detected. In the next detection, the already detected symbol does not participate in the new cost function computation. Therefore, the $l_0$-norm cost functions of each phase, are respectively given by
%\vspace{-5pt}
    \begin{eqnarray}\label{eq:cost_func_rls_l0}
    \hspace{-20pt}\mathcal{J}_{j}^\varphi\t\hspace{-7pt} &=&\hspace{-4.5pt}  \sum_{l=0}^{t} \hspace{5pt} \lambda^{t-l} \left|\varphi_{j}\left[l\right]-\mathbf{w}_j^\text{H}\t\mathbf{y}_{\psi_n}\t\right|^2 \hspace{-5pt}+ \gamma\|\mathbf{w}_j\t\|_0, \\ \label{eq:cost_func_rls2}
    \hspace{-20pt}\mathcal{J}_{j}^d\t\hspace{-7pt} &=&\hspace{-14pt} \sum_{l=\tau_{\phi}+1}^{t-1}\hspace{-7pt} \lambda^{t-1-l} \left|\hat{d}_{j}\left[l\right]-\mathbf{w}_j^\text{H}\t\mathbf{y}_{\psi_n}\t\right|^2 \hspace{-5pt}+ \gamma\|\mathbf{w}_j\t\|_0
    \end{eqnarray}

\noindent where $\|\cdot\|_0$ denotes $l_0$-norm that counts the number of zero entries in $\mathbf{w}_{j}$ and $\gamma$ is a non-zero positive constant to balance the regularization and, consequently, the estimation error. Moreover, $0< \lambda \leq 1$ is the forgetting factor which gives exponentially less weight to older error samples. $\mathcal{J}_{j}^\varphi\t$ represents in~(\ref{eq:cost_func_rls_l0}) the cost function for the metadata mode and $\mathcal{J}_{j}^d\t$ for the data mode.

\begin{figure*}[t]
        %% Conventional
        \subfloat[\footnotesize The internal list results are the continuous line while the external list values are shown in dashed lines. The probability of being active of each device is randomly drawn from a beta distribution with $\alpha=4$ and $\beta=8$.]{
        \includegraphics[scale=0.8]{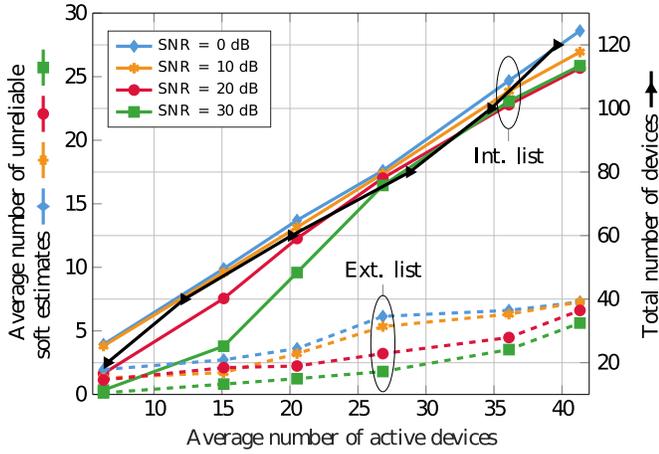}
        \label{fig:act_dev}
    }\hspace{50pt}
    \subfloat[The radius $r^{\text{th}_0}$ and $r^\text{th}$ delimits the reliable regions for the internal list while $r_\text{ext}$ for the external list.]{
    \centering
        \includegraphics[scale=.825]{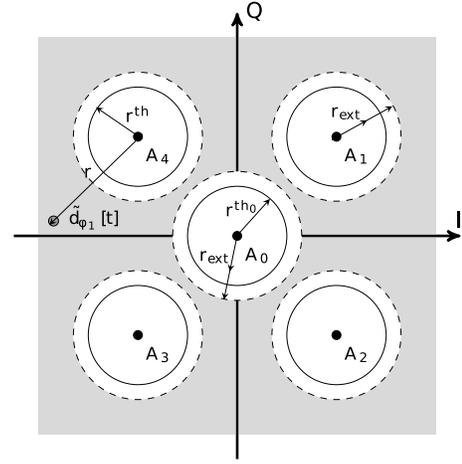}
        \label{fig:qpsk_aug}
    }
    \caption{\footnotesize Parameters of internal and external lists. (a) Number of unreliable soft estimates by the number of active devices, judge by the SAC of the internal and external list scheme and (b) Shadow area constraints for internal and external lists.}
    \label{fig:mmtc_scenario}
\end{figure*}

After the detection of all symbols, the vector with the soft estimates can be reorganized to the original order and integrate the iterative LDPC decoder. However, as for the next received vector AA-VGL-DF will verify which filter should be used first, we included another list detection technique at this point, in order to provide more refined information to the first cost function of the next received vector.

\subsubsection{External list}

The idea of the external list is to carry out a group list verification of the most unreliable symbols, after the last detection. In the internal list block, AA-VGL-DF also keeps the information about the reliability of each soft estimate. The $N \times 1$ binary vector $\boldsymbol{\vartheta}_\psi\t$ gathers this information as $\vartheta_{\psi_n}\t = 1$ when a soft estimate falls into the shadow area and $\vartheta_{\psi_n}\t =0$, otherwise. After the reordering process, the external list block receives the estimated symbols $\tilde{\mathbf{d}}\t$, the detected symbols $\hat{\mathbf{d}}\t$ and the vector with the reliability information, $\boldsymbol{\vartheta}\t$.

With the knowledge of which symbols had a reliable soft estimate, the external list generates all possible combinations $\mathcal{G}$ of the symbols of the considered augmented alphabet $\mathcal{A}_0$ and gathers all these vectors in an $\nu \times \mathcal{G}$ matrix $\mathbf{G}$ given by
%\vspace{-5pt}
\begin{equation}
    \mathbf{G} = \left[
    \begin{array}{ccccccc}
        \mathcal{A}_{0_{1}} & \mathcal{A}_{0_{1}} & \mathcal{A}_{0_{1}} & \cdots & \mathcal{A}_{0_{3}} & \cdots & \mathcal{A}_{0_{\left|\boldsymbol{\mathcal{A}}_{0}\right|}} \\
        \mathcal{A}_{0_{1}} & \mathcal{A}_{0_{2}} & \mathcal{A}_{0_{1}} & \cdots & \mathcal{A}_{0_{1}} & \cdots & \mathcal{A}_{0_{\left|\boldsymbol{\mathcal{A}}_{0}\right|}} \\
        \mathcal{A}_{0_{1}} & \mathcal{A}_{0_{1}} & \mathcal{A}_{0_{2}} & \cdots & \mathcal{A}_{0_{2}} & \cdots & \mathcal{A}_{0_{\left|\boldsymbol{\mathcal{A}}_{0}\right|}} \\
        \vdots & \vdots & \vdots & \ddots & \vdots & \ddots & \vdots \\
        \mathcal{A}_{0_{1}} & \mathcal{A}_{0_{1}} & \mathcal{A}_{0_{1}} & \cdots & \mathcal{A}_{0_{4}} & \cdots& \mathcal{A}_{0_{\left|\boldsymbol{\mathcal{A}}_{0}\right|}}
    \end{array}
    \right],
\end{equation}
\noindent where $\nu$ is the number of soft estimates considered unreliable. With the complete candidate vector matrix, the verification of the most appropriate vector occurs as in the internal list, as follows:
%\vspace{-5pt}
\begin{equation}
        \mathbf{g}_{\text{opt}} = \underset{i\,\in\, 1,\cdots,\mathcal{G}}{\textrm{arg min}}\hspace*{5pt} \left|\left|\,\mathbf{y}\t - \sum_ {j=1}^{\nu}\hat{\mathbf{h}}_{j}\, g_{j,i}\, \right|\right|^2,
\end{equation}

\noindent where $\hat{\mathbf{h}}_{j}$ is the estimated channel of the unreliable symbol to be verified. The vector candidate $\mathbf{g}_{\text{opt}}$ is chosen and its values replace the ones considered unreliable in $\hat{\mathbf{d}}\t$ in order to proceed to the detection, ordering and parameter estimation of the next received vector.

As an example, let us suppose that we have $4$ unreliable estimates (non-zero elements) in vector $\boldsymbol{\vartheta}_\psi\t$. Considering a QPSK modulation, the augmented alphabet $\mathcal{A}_0$ would have $5$ elements and $\mathcal{G}$ would be a $4 \times 70$ matrix, considering all the $70$ combination possibilities of $5$ possible symbols in $4$ unreliable estimates.

Since mMTC is a crowded scenario, building the external $\mathbf{G}$
matrix with all possible combinations would be impractical. Thus,
considering the distribution of probability of being active of
devices, we notice in Fig.~\ref{fig:act_dev} that just a few symbols
are considered unreliable by the internal list per received vector
and this number is reduced as the SNR grows. Therefore, we
considered a constraint in order to reduce the computational
complexity of the external list. Instead of verifying all the
possibilities of all unreliable soft estimates, we check the worst
cases, that is, the soft estimates considered more unreliable. Thus,
we define another radius in the SAC in order to designate which
unreliable symbols should be or not in the external list. As the
number of unreliable estimates varies with the SNR, we choose the
radius, represented in Fig.~\ref{fig:qpsk_aug}, as $r_\text{ext} =
r^{\text{th}}\left[\left(M/\hat{K}\right) + \left(N
\sigma_x^2/\sigma^2_v\right)\right]$ since it follows the increase
of the average SNR value. Thus, the reliability of
$\tilde{\mathbf{d}}\t$ will be rechecked, with the new radius
$r_\text{ext}$. As we have a $\hat{\mathbf{d}}\t$, we have an
estimation of which device is active or not, given by $\hat{K}$. So,
the number $\nu$ of considered unreliable soft estimates is reduced
as the SNR value grows, as shown in Fig.~\ref{fig:act_dev}. In the
next subsection, we detail the proposed $l_0$-norm regularized RLS
algorithm. Other list detection strategies can also be considered
\cite{spa,mfsic}.

%%%%%%%%%%%%%%%%%%%%%%%%%%%%%%%%%%%%%%%
\subsection{$l_0$-norm Regularized RLS Algorithm}

In order to exploit the sparse activity of devices and compute the parameters of the proposed DF detector without the need for explicit channel estimation, we devise an $l_0$-norm regularized RLS algorithm that minimizes the cost function. Approximating the value of the $l_0$-norm~\cite{Bradley98}, the cost function in (\ref{eq:cost_func_rls_l0}) can be rewritten as
%\vspace{-2.5pt}
    \begin{eqnarray}\nonumber
    \mathcal{J}_{j}^\varphi\t &=& \sum_{l=0}^{t} \lambda^{t-l} \left|\varphi_{j}\left[l\right]-\mathbf{w}_{j}^\text{H}\t\mathbf{y}_{\psi_n}\t\right|^2\\ \label{eq:cost_func_rls_l0_app}
     && + \gamma \sum^{2M}_{p=1} \left(1-\text{exp}\left(-\xi|w_{j,p}\t|\right)\right),
    \end{eqnarray}

\noindent where the parameter $\xi$ regulates the range of the attraction to zero on small coefficients of the filter. Thus, taking the partial derivatives for all entries $t$ of the coefficient vector $\mathbf{w}_{j}\t$ in (\ref{eq:cost_func_rls_l0_app}) and setting the results to zero, yields
%\vspace{-0.5pt}
    \begin{eqnarray}\nonumber
        \mathbf{w}_{j}\t &=& \mathbf{w}_{j}\left[t-1\right] + \mathbf{k}\t \epsilon_n^\ast\t \\ \label{eq:recursive_rls_l0}
        &&  - \gamma \, \xi\, \text{sgn}\left(w_{j,p}\t\right)\text{exp}\left(-\xi|w_{j,p}\t|\right)% e^{-\xi|w_{n,j}\t|}
    \end{eqnarray}

\noindent where $\mathbf{k}\t$ is the gain vector and $\text{sgn}\left(\cdot\right)$ is a component-wise sign function defined as
%   \vspace{-0.5pt}
    \begin{equation}
        \text{sgn}\left(w_{j,p}\t\right) =
        \left\{
        \begin{array}{rl}
%           \frac{w_{j,p}\t}{|w_{j,p}\t|}, & w_{j,p}\t \neq 0; \\
%           \nicefrac{w_{j,p}\t}{|w}_{j,p}\t|}, & w_{j,p}}\t \neq 0 \\
            w_{j,p}\t/|w_{j,p}\t|, & w_{j,p}\t \neq 0; \\
%           \left(w_{j,p}\t\right)/\left(|w_{j,p}\t|\right), & w_{j,p}\t \neq 0; \\
%           w_{j,p}\t/\left(|w_{j,p}\t|\right), & w_{j,p}\t \neq 0; \\
            0,                           & \text{otherwise.}
        \end{array}\right.
    \end{equation}

    In order to reduce computational complexity in (\ref{eq:recursive_rls_l0}), the exponential function is approximated by the first order of the Taylor series expansion, given by
%\vspace{-0.5pt}
    \begin{equation}\label{eq:exp_app}
        \text{exp}\left(-\xi\, |w_{j,p}\t|\right) \approx
        \left\{
        \begin{array}{rl}
            1-\xi\, |w_{j,p}\t|, & |w_{j,p}\t| \leq 1/\xi; \\
            0,                           & \text{otherwise.}
        \end{array}\right.
    \end{equation}

    As the exponential function is positive, the approximation of (\ref{eq:exp_app}) is also positive. In this way, (\ref{eq:recursive_rls_l0}) becomes
%\vspace{-2.5pt}
    \begin{eqnarray}\nonumber
        \mathbf{w}_{j}\t &=& \mathbf{w}_{j}\left[t-1\right] + \mathbf{k}\t \epsilon_n^\ast\t \\ \label{eq:recursive_rls_l0_app}
        &&  - \gamma \, \xi\, \text{sgn}\left(w_{j,p}\t\right)f_\xi\left(w_{j,p}\t\right)
    \end{eqnarray}

\noindent where the function $f_\xi\left(w_{j,p}\t\right)$ is given by

    \begin{equation}\label{eq:f_beta3}
        f_\xi\left(w_{j,p}\t\right) =
        \left\{
        \begin{array}{rl}
            \xi^2 \left(w_{j,p}\t\right)+\xi, & -1/\xi \leq w
            _{j,p}\t< 0; \\
            \xi^2 \left(w_{j,p}\t\right)-\xi, & 0 \leq w_{j,p}\t\leq 1/\xi; \\
            0,                           & \text{otherwise.}
        \end{array}\right.
    \end{equation}

We notice that the function $f_\xi\left(w_{j,p}\t\right)$ in
(\ref{eq:recursive_rls_l0_app}) imposes an attraction to zero of
small coefficients. So, if the value of $w_{j,p}\t$ is not equal or
in the range $\left[-1/\xi,1/\xi\right]$, no additional attraction
is exerted. Thus, the convergence rate of near-zero coefficients of
parameters of devices in mMTC applications that exhibit sparsity
will be increased~\cite{Bradley98}. The pseudo-code, which also
considers an IDD scheme with AA-VGL-DF, is described in Algorithm 1.
Alternatively, a designer can employ other more sophisticated
adaptive algorithms for parameter estimation and resource allocation
\cite{jio,jidf,jiols,jiomimo},
\cite{smce,1bitce,TongW,jpais_iet,armo,badstc,baplnc,goldstein,qian,jio,jidf,jiols,jiomimo,dce,jpba},
\cite{mwc,mwc_wsa,mwc_wsa,tds_cl,tds,armo,badstbc}.

%%%%%%%%%%%%%%%%%%%%%%%%%%%%%%%%%%%%%%%%%%%%%%%%%%%%%%%%%%%%%%%%%%%%%%%%%%%%%%%%%
% PROPOSED ITERATIVE DECODING
%%%%%%%%%%%%%%%%%%%%%%%%%%%%%%%%%%%%%%%%%%%%%%%%%%%%%%%%%%%%%%%%%%%%%%%%%%%%%%%%%
\section{Proposed Soft Information Processing and Decoding} \label{sec:prop_it_det_dec}

In order to devise an IDD scheme, we incorporate the detected
symbols by AA-VGL-DF in an iterative soft information decoding
scheme. Unlike existing approaches such as~\cite{AUchoa2016}, we
incorporate the probability of each device being active in the mMTC
scenario in the computation of each a priori probability symbol,
which avoids the need for channel estimation. Additionally, a
designer might consider preprocessing techniques
\cite{lclattice,switch_int,switch_mc,gbd,wlbd,mbthp,rmbthp,bbprec,1bitcpm,bdrs,baplnc,memd,wljio,locsme,okspme,lrcc}.

The \textit{a priori} probabilities are computed based on the
extrinsic LLRs $L_{e_n}^z\t$, provided by the LDPC decoder. In the
first iteration, all $L_{e_n}^z\t$ are zero and, assuming the bits
are statistically independent of one another, the \textit{a priori}
probabilities are calculated as
%\vspace{-5pt}
    \begin{equation}\label{eq:a_priori1}
        \text{Pr}\left(x_n\t = \overline{x}\right) = \sum_{\overline{x}\in \mathcal{A}_0} \overline{x}\left(\prod^{M_c}_{z=1}\left[1+\text{exp}\left(-\overline{x}^z L_{e_n}^z\t\right)\right]^{-1}\right),
    \end{equation}

\noindent where $M_c$ represents the total number of bits of symbol $\overline{x}$, the superscript $z$ indicates the $z$-th bit of symbol of $\overline{x}$, in $\overline{x}^z$ (whose value is $(+1,-1)$). As each device has a different activity probability $\rho_n$, the \textit{a priori} probabilities should take into account, as
%\vspace{-6.5pt}
\begin{align}
    %\left\{
    \begin{array}{rl}
     \hspace{-4pt}\rho_n + \left(1-\rho_n\right)\text{Pr}\left(x_n\t = \overline{x}\right), & \text{if}\, \left(\overline{x}^1 \text{ and } \overline{x}^2\right) = 0,    \\ \label{eq:a_priori2}
      \left(1-\rho_n\right)\text{Pr}\left(x_n\t = \overline{x}\right), & \text{otherwise}.
    \end{array}%\right.
\end{align}

%\vspace{-1pt}
\noindent where in the next iteration of the scheme, the new \textit{a priori} probabilities incorporates the probability that the $n$-th device is active and the extrinsic LLR values. As described in Section~\ref{sec:syst_model}, the probabilities $\rho_n$ are randomly drawn from a beta distribution.

As the output of the proposed receive filter has a large number of independent variables, we can approximate it as a Gaussian distribution~\cite{XWang1999}. Hence, we approximate $\tilde{d}_{n}\t$ by the output of an equivalent AWGN channel with $\tilde{d}_{n}\t = \mu_n\left[i\right] x_n\t + b_n\t$. Therefore, the likelihood function $P\left(\tilde{d}_{n}\t|\,\overline{x}\right)$ is approximated by
%\vspace{-5pt}
    \begin{equation}\label{eq:likelihood}
        P\left(\tilde{d}_{n}\t|\, \overline{x}\right) \approx \frac{1}{\pi\, \zeta^2_n\t} \text{exp}\left(-\frac{1}{\zeta_n^2\t}|\tilde{d}_{n}\t-\mu_n\t \overline{x}|^2\right),
    \end{equation}

\noindent where the mean $\mu_n\t$ is given by
%\vspace{-1pt}
    \begin{eqnarray}\label{eq:mean_gauss_app} \nonumber
        \mu_n\t &=& \mathbb{E}\left\{\tilde{d}_{n}\t x_{n}\t\right\}= \mathbb{E}\left\{\mathbf{w}_n^\text{H}\t\mathbf{y}_n\t x_{n}\t\right\} \\
        &\approx & \mathbf{w}_n^\text{H}\t \left(\sum_{p=1}^{t-1}\lambda^{t-1-p}\, \mathbf{y}_n\left[p\right]x_{n}\left[p\right]\right). \\ \nonumber
    \end{eqnarray}
\vspace{-1pt}
\noindent Note that $x_n$ is the previously detected symbol. In the first case, $\mu_n\t = \mathbf{w}_n^\text{H}\t \mathbf{y}_n\t$. Each $b_n\t$ is a zero-mean complex Gaussian variable with variance $\zeta^2_n\t$ as
%\vspace{-1pt}
    \begin{eqnarray}\label{eq:var_gauss_app} \nonumber
        \zeta_n^2\t &=& \text{var}\left\{\tilde{d}_{n}\t\right\} = \mathbb{E}\left\{\|\tilde{d}_{n}\t\|^2\right\}-\mu_n^2\t  \\
        &=& \mathbf{w}_n^\text{H}\t\mathbb{E}\left\{{\mathbf{y}_n}\t\mathbf{y}_n^\text{H}\t\right\}\mathbf{w}_n\t - \mu_n^2\t \\ \nonumber
        &\approx& \mathbf{w}_n^\text{H}\t\left(\sum_{p=1}^{t}\lambda^{t-p}\, \mathbf{y}_n\left[p\right]\mathbf{y}_n^\text{H}\left[p\right]\right)\mathbf{w}_n\t - \mu_n^2\left[i\right].
    \end{eqnarray}

Then, the extrinsic LLRs computed by the AA-VGL-DF detector for the $z$-th bit ($z \in \{1,\dots,M_c\}$) of the symbol $x_n$ transmitted by the $n$-th device are given by
%\vspace{-1pt}
\begin{align}\label{eq:llr}
        L_{c_n}^z\t =& \log \frac{\sum_{\overline{x}\in \mathcal{A}_z^{+1}}\text{Pr}\left(\tilde{d}_{n}\t|\, \overline{x}\right)\text{Pr}\left(\overline{x}\right)}{\sum_{\overline{x}\in \mathcal{A}_z^{-1}}\text{Pr}\left(\tilde{d}_{n}\t|\, \overline{x}\right)\text{Pr}\left(\overline{x}\right)} - L_{e_n}^z\t
\raisetag{20pt}
\end{align}%}}

\noindent where $\mathcal{A}_z^{+1}$ is the set of $2^{M_c -1}$ hypotheses of $\overline{x}$ for which the $z$-th bit is +1 (analogously for $\mathcal{A}_z^{-1}$).

%%%%%%%%%%%%%%%%%%%%%%%%%%%%%%%%%%%%%%%%%%%%%%%%%%%%%%%%%%%%%%%%%%%%%%%%%%%%%%%%%
% ANALYSIS OF THE AA-VGL-DF ALGORITHM
%%%%%%%%%%%%%%%%%%%%%%%%%%%%%%%%%%%%%%%%%%%%%%%%%%%%%%%%%%%%%%%%%%%%%%%%%%%%%%%%%
\section{Analysis of the AA-VGL-DF Algorithm} \label{sec:analysis}
In this section, the computational complexity required by the AA-VGL-DF algorithm is evaluated and both the diversity order achieved by the AA-VGL-DF detector and the achievable rate of the uplink transmission from the $n$-th user are discussed.

% Algorithm pseudocode
% -------------------------------------------------------------------------
\begin{table}[t]
%%\normalsize
%\small
    \begin{center}
        \begin{tabular}{p{0.2cm}l} \\ \hline
        \multicolumn{2}{l}{\textbf{Algorithm 1} Proposed IDD with AA-VGL-DF} \\ \hline
            1. & Initialization: $M$, $N$, $\bm{\rho}$, $\xi$, $\gamma$, $\lambda$, $\mathbf{P}_{\psi_n} = \bm{\rho}\,\mathbf{I}_{\text{M}}$\\[2pt]
               & \footnotesize\textbf{ \% For training mode,} \\[2pt]
               & \footnotesize \hspace{10pt} \% For each metadata sequence $\hat{\mathbf{d}}\t$ and $\mathbf{y}_{{\psi_n}}\t$, \\[2pt]
            \small
            2. & \hspace{10pt} Compute the Kalman gain vector\\[2pt]
               & \hspace{10pt} $\scriptstyle \mathbf{k}_{\psi_n}\t = (\mathbf{P}_{\psi_n}\t \,\mathbf{y}_{{\psi_n}}\t)/(\lambda+\mathbf{y}_{{\psi_n}}^\text{H}\t\, \mathbf{P}_{\psi_n}\t \, \mathbf{y}_{{\psi_n}}\t)$;\\[2pt]
            3. & \hspace{10pt} Estimate $\tilde{d}_{\psi_n}\t = \mathbf{w}_n^\text{H}\t\,\mathbf{y}_{n}\t$;\\[2pt]
            4. & \hspace{10pt} Update the error value with $\epsilon_{\psi_n}\t = \hat{d}_{\psi_n}\t - \tilde{d}_{\psi_n}\t$;\\[2pt]
            5. & \hspace{10pt} Update the filters with Eq.~(\ref{eq:recursive_rls_l0_app});\\[2pt]
            6. & \hspace{10pt} Update the auxiliary matrix\\[2pt]
               & \hspace{10pt} $\mathbf{P}_{\psi_n}\t = \lambda^{-1} \left(\mathbf{P}_{\psi_n}\t - \mathbf{k}_{\psi_n}\t\, \mathbf{y}_{{\psi_n}}^\text{H}\t\,\mathbf{P}_n\t\right)$; \\[2pt]
            7. & \hspace{10pt} Concatenate $\mathbf{y}_{\psi_n}\t$ with $\hat{d}_{\psi_{n}}\t$; \\[2pt]
            8. & \hspace{10pt} Update the sequence of detection with Eq.(\ref{eq:detec_ord}); \\[2pt]
               & \footnotesize \textbf{\% For decision-directed mode,}\\[2pt]
               \small
            9. & \hspace{10pt} Compute the \textit{a priori probability} with Eqs.~(\ref{eq:a_priori1}) and (\ref{eq:a_priori2}); \\[2pt]
           10. & \hspace{10pt} Repeat steps $2.$ to $6.$; \\[2pt]
           11. & \footnotesize \hspace{10pt} Evaluate the reliability of the soft estimation $\tilde{d}_{\psi_n}\t$ with SAC \\[2pt]
               & \footnotesize \hspace{10pt} and proceeds with the internal list if it is judged as unreliable; \\[2pt]
           12. & \hspace{10pt} Update the sequence of detection with the output of 11; \\[2pt]
           13. & \footnotesize \hspace{10pt} Proceed with the update of $\bm{\vartheta}_{\psi}\t$, $\mathbf{y}_{\psi_n}\t$ and $\mathbf{w}_{\psi_n}\t$;\\[2pt]
           14. & \hspace{10pt} After all detections, update $\hat{\mathbf{d}}\t$ with the external list;\\[2pt]
           \small
           15. & \hspace{10pt} Compute $\mu_{\psi_n}\t$ and $\zeta^2_{\psi_n}\t$ with Eqs.~(\ref{eq:mean_gauss_app}) and (\ref{eq:var_gauss_app});\\[2pt]
           16.  & \hspace{10pt} Verify the likelihood function $P\left(\tilde{d}_{n}\t|\overline{x}\right)$ with Eq.~(\ref{eq:likelihood}); \\[2pt]
           17.  & \hspace{10pt} Compute the LLR value according to Eq.(\ref{eq:llr}). \\ \hline
        \end{tabular}
    \end{center}
\end{table}

% Computational Complexity
% -------------------------------------------------------------------------
\subsection{Computational Complexity}\label{subsec:compl}

The computational complexity of AA-VGL-DF is analyzed below by counting each required numerical operation in terms of complex FLOPs. In particular, Table~\ref{tab:complexity} compares the number of required FLOPs, for a different number of devices $N$, receive antennas $M$ and the $\mathcal{G}$ group size. We consider both well-known algorithms as linear minimum-mean-squared-error (LMMSE) and modifications for mMTC, as SA-SIC~\cite{Knoop2013}, SA-SIC with A-SQRD\cite{Ahn2018} and AA-RLS-DF~\cite{DiRennaWCL2019}.

\begin{table*}[t]
    \centering
    \caption{\footnotesize FLOPs counting of considered techniques in detail.}
    \label{tab:complexity}
    \begin{tabular}{lc} \hline
        Algorithms                  & Required number of FLOPs \\ \hline
        LMMSE                       & $2M^3 + 4\left(N+1\right)M^2 + 2\left(N^2+N+1\right)M -\left(N^2+N\right)$ \\
        SA-SIC \cite{Knoop2013} & $|\mathcal{A}_0| \left(N^3 + N^2 + 6\right)$ \\
        SA-SIC A-SQRD \cite{Ahn2018} & $2N^3 + 4\left(M +1\right)N^2 +\left(M-1\right)N$ \\
        AA-RLS Linear               & $\left(6M^2 + 10M\right)N$ \\
        AA-RLS Linear (internal list)   & $\left[6M^2 + 10M + \vartheta_{n}\left(2M|\mathcal{A}_0|\right)\right]N$ \\
        AA-RLS-DF \cite{DiRennaWCL2019} & $\sum_{i=1}^{N} \left[6\left(M+i\right)^2 + 10\left(M+i\right)\right]$ \\
        AA-RLS-DF (internal list) & $\sum_{i=1}^{N} \left[6\left(M+i\right)^2 + 10\left(M+i\right)+ \vartheta_{n}\left(2M|\mathcal{A}_0|\right)\right]$\\
        AA-VGL-DF                   & $\sum_{i=1}^{N} \left[6\left(M+i\right)^2 + 10\left(M+i\right)+ \vartheta_{n}\left(2M|\mathcal{A}_0|\right)\right]\, + 2M\mathcal{G}$ \\
        \hline
    \end{tabular}
\end{table*}

\begin{figure}[t]
    \centering
    \includegraphics[scale=0.97]{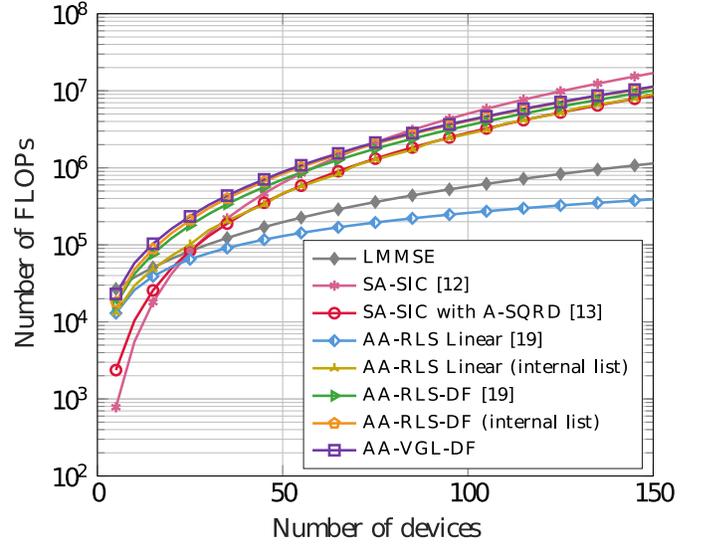}
    \caption{\footnotesize Comparison of complexity of considered algorithms. The values chosen were $M=20$, $\tau_{\phi}=N/2$ and the variables related to the lists $\mathcal{G}$ reach the maximum of 5. Just 25\% of $\boldsymbol{\vartheta}$ is equal to 1, following the beta-binomial distribution as $\alpha = 4$ and $\beta = 8$.}
    \label{fig:complexity}
\end{figure}

Including the internal list scheme in our previous work~\cite{DiRennaWCL2019}, AA-RLS-DF, as shown in Fig.~\ref{fig:complexity}, results in just a slight complexity increase. Recalling that $\mathcal{G}$ is the number of combinations of the unreliable soft estimates, the upper bound of AA-VGL-DF is a version where there are not constraints in the lists. As verified in Fig.~\ref{fig:act_dev}, the number of unreliable soft estimates increase as the number of devices rises. The computational cost of AA-VGL-DF is comparable with a standard DF detector with an RLS algorithm. Since many other considered algorithms have a similar computational cost, AA-VGL-DF has a competitive complexity when compared with other schemes.

% Uplink sum-rate
% -------------------------------------------------------------------------
\subsection{Uplink Sum-Rate}\label{subsec:ul}

As seen that the mMTC has an amount of features that distinguish from the standard massive MIMO communications, we compute the uplink sum-rate considering our detector. Whereas the scenario implies the consideration of a different number of active devices at the same transmission slot and the probability of collision due to reuse of metadata sequences, we compute the achievable rate of each device. We take into account all the possible contamination events from the active devices, as the number of transmission slots are large enough. Were also considered that the BS can estimate the number of active devices, as well as the average channel energy $\eta_n$. In this way, each device has the knowledge of its channel energy and is able to associate it with its rate, as both parameters are broadcast by the BS.

Differently from the literature, the expressions derived take into account, beyond the filter computed by each approach, the probability of metadata collisions, the probability of having an specific number of active devices and different features of each device, as variable activity probability, transmission power and the path loss and shadowing experienced.

\textit{Theorem 1:} An approximation of a lower bound of the maximal achievable sum-rate (in bits per symbol) is
%\vspace{-1.5pt}
\begin{equation}
    R = \sum_{K=1}^{N} p\left(K\right)K \sum_{c=0}^{K-1} p\left(c|K\right) \mathbb{E}_{\left\{\eta\right\}} \left[\mathcal{R}\left(\mathcal{C}_i,K,\left\{\eta\right\}\right)\right]
\end{equation}

\noindent where $p\left(K\right)$, given by~(\ref{eq:pdf}), is the probability of having $K$ active devices in a total of $N$ and $p\left(c|K\right)$ is the probability of having $c$ devices with the same metadata sequence of the $i$-th device being observed out of $K$ active devices and $\mathcal{C}_i$ refers to a set of those contaminator devices. The expression of $p\left(c|K\right)$ is given by

\begin{equation}\label{eq:prob_col}
    p\left(c|K\right) = \begin{pmatrix} K-1 \\ c \end{pmatrix} \left(\frac{1}{\tau_\phi}\right)^c \left(1-\frac{1}{\tau_\phi}\right)^{K-1-c}.
\end{equation}

The procedure and the derivation of the first summations are given by~\cite{ECarvalho2017}. $\mathbb{E}_{\left\{\eta\right\}}$ designates the expectation with respect to $\eta_j, j \in \left\{i,\mathcal{C}_i\right\}$ and $\mathcal{R}\left(\mathcal{C}_i,K,\left\{\eta\right\}\right)$ is a lower bound on the maximal achievable rate of device $i$ conditioned on a collider set with indices $\mathcal{C}_i$ within $K$ active devices is given by
%\vspace{-1.5pt}
\begin{equation}
    \mathcal{R}\left(\mathcal{C}_i,K,\left\{\eta\right\}\right) = \log_2\left(1+ \text{SINR}\left(\mathcal{C}_i,K,\left\{\eta\right\}\right)\right).
\end{equation}

\subsubsection{Perfect Channel Estimation}

We first consider the case when the BS has perfect CSI, i.e., it has the perfect knowledge of $\mathbf{H}$. Therefore, the channel capacity $C$, recalling $\tilde{\mathbf{d}}\t$ and $\mathbf{y}\t$ in~(\ref{eq:out_equa}) and suppressing the time index for simplicity, is
%\vspace{-1.5pt}
\begin{align}
    C    &= \underset{p_{\tilde{\mathbf{d}}}\left(\tilde{\mathbf{d}}\right)}{\textrm{max}}      \hspace*{2pt} I\left(\mathbf{y};\tilde{\mathbf{d}}\right) \\ \nonumber
         &= \underset{p_{\tilde{\mathbf{d}}}\left(\tilde{\mathbf{d}}\right)}{\textrm{max}}      \hspace*{2pt} H\left(\mathbf{y}\right) - H\left(\tilde{\mathbf{d}}|\mathbf{y}\right) = \underset{p_{\tilde{\mathbf{d}}}\left(\tilde{\mathbf{d}}\right)}{\textrm{max}}   \hspace*{2pt} H\left(\mathbf{W}^\text{H}\mathbf{y}\right) - H\left(\mathbf{W}^\text{H}\mathbf{v}\right)
\end{align}

\noindent where $H$ is the differential entropy and $I$ the mutual information. Thus, as the considered signals are Gaussian, the mutual information is given by~\cite{Tse2005}
%\vspace{-1.5pt}
\begin{align} \label{eq:mut_inf}
    &\hspace{-5pt}I\left(\mathbf{y};\tilde{\mathbf{d}}\right) =\\ \nonumber
    &\hspace{-5pt}\log_2 \left(\text{det}\left(\mathbb{E}\left[\mathbf{W}^\text{H}\,\mathbf{y}\mathbf{y}\,^\text{H}\mathbf{W}\right]\right)\right) - \log_2 \left(\text{det}\left(\mathbb{E}\left[\mathbf{W}^\text{H}\,\mathbf{v}\mathbf{v}\,^\text{H}\mathbf{W}\right]\right)\right).
\end{align}

Thus, computing $\Omega = \mathbb{E}\left[\mathbf{W}^\text{H}\,\mathbf{y}\mathbf{y}\,^\text{H}\mathbf{W}\right]$, we have to rewrite the following:
%\vspace{-2.5pt}
\begin{align}
    &\Omega = \mathbb{E}\left[\sum_{j=1}^{K}\mathbf{w}_j^\text{H}\,y_j y_j\,^\text{H}\mathbf{w}_j\right]  \\ \nonumber
    &\hspace{7pt}= \mathbb{E}\left[\sum_{j=1}^{K}\mathbf{w}_j^\text{H}\,\left(\mathbf{h}_j\sqrt{b_j}\,x_j+\mathbf{v}\right)\left(\mathbf{h}_j\sqrt{b_j}\,x_j+\mathbf{v}\right)^\text{H}\mathbf{w}_j\right] \\ \nonumber
    &\hspace{7pt}= \mathbb{E}\left[\sum_{j=1}^{K}\,\left(\mathbf{w}_j^\text{H}\,\mathbf{h}_j\sqrt{b_j}\,x_jx_j^\text{H}\sqrt{b_j}\mathbf{h}_j^\text{H}\,\mathbf{w}_j\right)+\right.\\ \nonumber
    & \hspace{77pt} \left(\mathbf{w}_j^\text{H}\,\mathbf{h}_j\sqrt{b_j}\,x_jv_j^\text{H}\mathbf{w}_j\right) +  \\ \nonumber
    & \hspace{5pt} \left. {\color{white}{\sum_{j=1}^{K}}} \hspace{55pt} \left(\mathbf{w}_j^\text{H}v_jx_j^\text{H}\sqrt{b_j}\mathbf{h}_j^\text{H}\,\mathbf{w}_j\right) + \left(\mathbf{w}_j^\text{H}v_j v_j^\text{H}\mathbf{w}_j\right) \right].
\end{align}

As the main objective is to compute the maximum achievable rate of a device $i$ out of $K$ active devices in the same time instant $t$ we have,
%\vspace{-5pt}
\begin{align} \nonumber
\Omega &= \mathbb{E}\left[\sum_{j=1}^{K}\,\left(\mathbf{w}_i^\text{H}\,\mathbf{h}_j\sqrt{b_j}\,x_jx_j^\text{H}\sqrt{b_j}\mathbf{h}_j^\text{H}\,\mathbf{w}_i\right)+ \left(\mathbf{w}_i^\text{H}\mathbf{v}\mathbf{v}^\text{H}\mathbf{w}_i\right) \right]\\  \nonumber
& = \mathbb{E}\left[\left|\mathbf{w}_i^\text{H}\,{\mathbf{h}}_i \sqrt{b_i}\,x_i\right|^2\hspace{-2.5pt} +\hspace{-5.5pt} \sum_{j=1,j\neq i}^{K}\hspace{-2pt}\left|\mathbf{w}_i^\text{H}\,{\mathbf{h}}_j \sqrt{b_j}\,x_j\right|^2 + \left|\mathbf{w}_i^\text{H}\mathbf{v}\right|^2\right] \\ \label{eq:deriv3}
\end{align}

\noindent where the first term is the signal of interest and the other additive terms are treated as a Gaussian noise. Thus, substituting~(\ref{eq:deriv3}) in~(\ref{eq:mut_inf}), we get the expression of the signal-to-noise-plus-interference for the fixed channel realization $\mathbf{H}$, as
%\vspace{-2.5pt}
\begin{align} \label{eq:mut_inf2}
&\text{SINR} = \frac{\left|\mathbf{w}_i^\text{H}\,{\mathbf{h}}_i \sqrt{b_i}\right|^2}{\sum\limits_{j=1,j\neq i}\hspace{-2.5pt}\left|\mathbf{w}_i^\text{H}\,{\mathbf{h}}_j \sqrt{b_j}\right|^2 + \|\mathbf{w}_i^\text{H}\|^2}.% = \frac{\mathbf{w}_i^\text{H}\,\eta_i \,b_i\, \mathbf{w}_i}{\mathbf{w}_i^\text{H}\,\sum\limits_{j=1,j\neq i}\hspace{-2.5pt}\eta_j b_j} .
\end{align}

%%%%%%% Imp CSI
\subsubsection{Imperfect Channel Estimation}\label{subsubsec:ImpCE}

In practice, the channel matrix $\mathbf{H}$ has to be estimated at the BS. Thus, we define the computation of the LMMSE estimate of the channel estimate of the $n$-th device, as
%\vspace{-1.5pt}
\begin{eqnarray} \nonumber
    \mathbf{y}_n &=& \mathbf{Y}_\phi \, \boldsymbol{\varphi}_n^H = \sum_{n' \in N} \left(\sqrt{\tau_\phi\, b_{n'}}\,\mathbf{h}_{n'}\,\boldsymbol{\varphi}_{n'} + \mathbf{V}_\phi\right)\boldsymbol{\varphi}_n^H, \\ \nonumber
    &=& \sqrt{\tau_\phi\, b_{n}}\,\mathbf{h}_n +\\
    && \sum_{n' \neq n} \sqrt{\tau_\phi\, b_{n'}}\,\mathbf{h}_{n'}\,\boldsymbol{\varphi}_{n'}\boldsymbol{\varphi}_{n}^H + \mathbf{V}_\phi\,\boldsymbol{\varphi}_n^H
\end{eqnarray}

\noindent where $\boldsymbol{\varphi}_n$ is the $1 \times \tau_\phi$ metadata vector of the $n$-th device, $\mathbf{V}_\phi$ is the $M \times \tau_\phi$ noise matrix and the components of $\left(\mathbf{V}_\phi\,\boldsymbol{\varphi}_n^H\right)$ are i.i.d., as $\|\boldsymbol{\varphi}_n\|^2$. Then, the LMMSE estimate of $\mathbf{h}_n$, $\hat{\mathbf{h}}_n$ is

\begin{eqnarray}\nonumber
    \hat{\mathbf{h}}_n &=& \frac{\mathbb{E}\left\{\mathbf{y}_n^H\mathbf{h}_n\right\}}{\mathbb{E}\left\{\mathbf{y}_n\mathbf{y}_n\right\}}\mathbf{y}_n, \\
    &=& \frac{\eta_n\sqrt{\tau_\phi\, b_{n}}}{\sum_{n'\in N} \tau_\phi\, b_{n'}\,\eta_{n'}\,|\boldsymbol{\varphi}_{n'}\boldsymbol{\varphi}_{n}^H|^2 + \,\sigma_v^2}\mathbf{y}_n.
\end{eqnarray}

Thus, $\hat{\mathbf{H}}$ is the $N \times M$ matrix of channel estimate. We denote $\bm{\mathcal{E}} = \hat{\mathbf{H}} - \mathbf{H}$, where the elements of $\bm{\mathcal{E}} = \left[{\scriptstyle{\bm{\mathcal{E}}}}_1,{\scriptstyle{\bm{\mathcal{E}}}}_2, \cdots, {\scriptstyle{\bm{\mathcal{E}}}}_M\right]$ are random variables with zero mean and variance $\left(\eta_i\right)/\left(b_i \eta_i +1\right)$. Furthermore, owing to the properties of LMMSE estimation, $\bm{\mathcal{E}}$ is independent of $\hat{\mathbf{H}}$. Splitting~(\ref{eq:deriv3}) in devices with and without the same metadata sequence assigned to the $i$-th device, we have
%\vspace{-5pt}
\begin{align} \label{eq:deriv4}
\Omega = &\hspace{2.5pt} \mathbb{E}\left[\sum_{j \in\left\{i,\mathcal{C}_i\right\}}\,\left(\mathbf{w}_i^\text{H}\,\mathbf{h}_j\sqrt{b_j}\,x_jx_j^\text{H}\sqrt{b_j}\mathbf{h}_j^\text{H}\,\mathbf{w}_i\right)\right.+\\ \nonumber
&\left. \sum_{j \notin\left\{i,\mathcal{C}_i\right\}}\,\left(\mathbf{w}_i^\text{H}\,\mathbf{h}_j\sqrt{b_j}\,x_jx_j^\text{H}\sqrt{b_j}\mathbf{h}_j^\text{H}\,\mathbf{w}_i\right) + \left(\mathbf{w}_i^\text{H}\mathbf{v}\mathbf{v}^\text{H}\mathbf{w}_i\right) \right],
\end{align}

Recalling that $\boldsymbol{\mathcal{E}} = \hat{\mathbf{H}}-\mathbf{H}$ and considering the independence between $\bm{\mathcal{E}}$, $\hat{\mathbf{H}}$ and separating the signal of interest we obtain,
%\vspace{-8pt}
\begin{align} \label{eq:deriv5}
\Omega = & \\ \nonumber
& \mathbb{E}\left[\left|\mathbf{w}_i^\text{H}\,\hat{\mathbf{h}}_i \sqrt{b_i}\,x_i\right|^2\hspace{-2.5pt} +\hspace{-3pt} \sum_{j \in\left\{\mathcal{C}_i\right\}}\,\left|\mathbf{w}_i^\text{H}\,\hat{\mathbf{h}}_j \sqrt{b_j}\,x_j\right|^2\hspace{-2.5pt} + \left|\mathbf{w}_i^\text{H}\mathbf{v}\right|^2+ \right.\\ \nonumber
& \hspace{10pt} \left. \sum_{j \in\left\{i,\mathcal{C}_i\right\}}\,\left|\mathbf{w}_i^\text{H}\,{\scriptstyle{\bm{\mathcal{E}}}}_j \sqrt{b_j}\,x_j\right|^2+ \sum_{j \notin\left\{i,\mathcal{C}_i\right\}}\,\left|\mathbf{w}_i^\text{H}\,\mathbf{h}_j\sqrt{b_j}\,x_j\right|^2 \right]
\end{align}

\noindent where the first term is the signal of interest and the other additive terms are treated as a Gaussian noise. Thus, substituting~(\ref{eq:deriv5}) in~(\ref{eq:mut_inf}), we get the expression,~(\ref{eq:mut_inf3}) in the top of the next page.

\begin{figure*}
% ensure that we have normalsize text
\normalsize
\setcounter{MYtempeqncnt}{\value{equation}}
% Set the equation number to one less than the one
% desired for the first equation here.
% The value here will have to changed if equations
% are added or removed prior to the place these
% equations are referenced in the main text.
\setcounter{equation}{44}
\begin{align} \nonumber
\text{SINR} =&\hspace{70pt}\frac{\left|\mathbf{w}_i^\text{H}\,\hat{\mathbf{h}}_i \sqrt{b_i}\right|^2\, }{\sum\limits_{j \in\left\{\mathcal{C}_i\right\}}\,\left|\mathbf{w}_i^\text{H}\,\hat{\mathbf{h}}_j \sqrt{b_j}\right|^2 + \sum\limits_{j \in\left\{j,\mathcal{C}_i\right\}}\,\left|\mathbf{w}_i^\text{H}\,{\scriptstyle{\bm{\mathcal{E}}}}_j \sqrt{b_j}\right|^2+ \sum\limits_{j \notin\left\{i,\mathcal{C}_i\right\}}\,\left|\mathbf{w}_i^\text{H}\,\mathbf{h}_j\sqrt{b_j}\right|^2 + \|\mathbf{w}_i^\text{H}\|^2} \\  \label{eq:mut_inf3}
 =&\hspace{10pt} \frac{\mathbf{w}_i^\text{H}{b_i}\left(\eta_i-\left(\left(b_j\,\eta_i^2\right)\bigg/\left(\sqrt{b_j}\eta_i + \sum\limits_{j \in C_i} \eta_j +1\right)\right)\right) \mathbf{w}_i}{\mathbf{w}_i^\text{H}\left(\sum\limits_{j \in\left\{\mathcal{C}_i\right\}}\,b_j\,\left(\eta_i-\frac{b_j\,\eta_i^2}{\sqrt{b_j}\eta_i + \sum\limits_{j \in C_i} \eta_j +1}\right) + \sum\limits_{j \in\left\{j,\mathcal{C}_i\right\}}\,b_j \left(\frac{\eta_j}{b_j\eta_j +1}\right) + \sum\limits_{j \notin\left\{i,\mathcal{C}_i\right\}}\,b_j\,\eta_j\right)\mathbf{w}_i + \left(\frac{\left(1-\lambda\right)^2\, \sigma_v^2\,\sigma_y^2 +1}{2\left(1-\lambda\right)}\right)}
\end{align}
% Restore the current equation number.
%\setcounter{equation}{\value{MYtempeqncnt}}
\setcounter{equation}{45}
% The IEEE uses as a separator
\hrulefill
% The spacer can be tweaked to stop underfull vboxes.
\vspace*{4pt}
\end{figure*}

% Diversity Order
% -------------------------------------------------------------------------
\subsection{Diversity Order}\label{subsec:div}
This section is devoted to present the diversity order achieved by the AA-VGL-DF detector. We adopt the geometrical approach presented in~\cite{SLoyka2004} and used in the previous work~\cite{deLamare2013} in order to reach the expression. As for non-ergodic scenarios the error probability is the probability that the signal level is less than the specified value, also known as outage probability, the diversity order, that is, the asymptotic slope of the outage probability curve~\cite{LZheng2003}\cite{HZhang2009}, is given by

\begin{equation}
    d \triangleq  \underset{x \rightarrow \infty}{\text{lim}}\hspace*{5pt} \frac{\log\left(P_r\left(R_{k,\text{span}\left\{\overline{k}\right\}}\right)\leq x\right)}{\log \left(x\right)}
\end{equation}

\noindent where $R_{k,\text{span}\left\{\overline{k}\right\}} = R_{k,\text{span}\left\{1,2,\dots,k-1,k+1,\dots,K\right\}}$ is the squared projection height from the $k$th column vector $\mathbf{h}_K$ of $\mathbf{H}$. From the definition in~\cite{HZhang2009}, $R_{k,\text{span}\left\{\overline{k}\right\}}= \|\bm{\Upsilon}\, \mathbf{h}_K\|^2$ where $\bm{\Upsilon} = \mathbf{I}-\mathbf{P}\mathbf{P}^\text{H}$ is the projection matrix to the orthogonal space of $\text{span}\left\{\overline{k}\right\}$ and $\mathbf{P}$ is composed of any orthonormal bases of this subspace. An important point is that only the $K$ active devices are considered for the computation of the diversity order.

\textit{Theorem 2: The diversity order achieved by the AA-VGL-DF detector is given by}
%\vspace{-1pt}
\begin{align}\label{eq:div_ord}
    d_\text{VGL} = & M - K +\left(\bm{\vartheta}^\text{T}\,\mathbf{m_\text{ord}} + \vartheta^0\right) + \mathcal{G}
\end{align}

\noindent \textit{where $\bm{\vartheta}$ is the $K \times 1$ binary vector presented in Section~\ref{subsec:adapt_imp} that gathers the information about the reliability of the soft estimates. $\mathbf{m_\text{ord}} = \left[|\mathcal{A}|,|\mathcal{A}|,\dots,|\mathcal{A}|\right]^\text{T}$ is also a $K \times 1$ vector but each column has the number of symbols of the considered alphabet and $\mathcal{G}$ is the number of all the possible combinations of the symbols of the considered augmented alphabet generated by the external list. $\vartheta^0$ is the total number of zeros in the vector $\bm{\vartheta}$.}

\textit{Proof:} As the decision feedback scheme applies an interference cancellation at each detection step, as it is common in the literature~\cite{MKVaranasi1999,AARontogiannis2006}, we can make an analogy to the well-known successive interference cancellation (SIC) scheme. Assuming the channel model described in Section~\ref{sec:syst_model} and making the common assumption that there is no error propagation related to the interference cancellation~\cite{deLamare2013,LZheng2003,HZhang2009,SLoyka2004}, the interference nulling out can be expressed as a general matrix form given by
%\vspace{-5pt}
    \begin{equation}\label{eq:proj}
        \mathbf{y}_\perp = \bm{\Upsilon} \cdot \mathbf{y},
    \end{equation}

\noindent where (\ref{eq:proj}) projects $\mathbf{y}$ onto the direction orthogonal to the span $\left\{\overline{k}\right\}$. Following the procedure in~\cite{SLoyka2004}, to reach the expression of the diversity order of each step of the SIC, the idea is to rotate the set of channels $\left[\mathbf{h}_1, \dots, \mathbf{h}_K\right]$ in a way that $\mathbf{h}_K$ becomes parallel to $\mathbf{p}_K$, one of the orthonormal basis of the subspace. Considering the detection of the first step, $\mathbf{h}_K$ is fixed and position $\mathbf{h}_{K-1}$ into the $\left[\mathbf{p}_{K-1}\, \mathbf{p}_K\right]$ plane. In this way, the received signal vector can be written as
%\vspace{-2.5pt}
    \begin{equation}
        \mathbf{y}_\perp = \sqrt{b_1 \tau_x}\, x_1 \bm{\Upsilon} \cdot \mathbf{h}_1 + \mathbf{v},
    \end{equation}

\noindent where the time instants are suppressed to reduce the notation. The rotations happens until the last channel vector, $\mathbf{h}_K$, is positioned into the $\left[\mathbf{p}_2\, \mathbf{p}_3, \dots, \mathbf{p}_K\right]$ hyper plane. In the well-known SIC scheme, the diversity order after all rotations is $\left(M-K+1\right)$, as $\mathbf{h}_{1\perp}$ has a total of $\left(M-K+1\right)$ nonzero components. On the other hand, as the AA-VGL-DF scheme has an internal list at each detection step, the diversity gain can be increased.

Assuming that the first soft estimate was considered unreliable by the SAC, the internal list scheme would imply more than one possible received vector to be cancelled. Thus, designating the order of the alphabet of the chosen modulation scheme as $|\mathcal{A}|$, for the first step, the diversity order is $\left(M-K+|\mathcal{A}|\right)$. For the steps that the soft estimation is reliable, the diversity order is the same as the SIC scheme. Thereby, the result can be achieved by induction. For the $i$th step, the diversity order can be represented by
%\vspace{-1pt}
\begin{equation}
\left\{ \begin{array}{ll}
        M - K + \vartheta_i^0, & \text{if }  \tilde{d}_{n} \text{ is reliable and} \\
        M - K + \bm{\vartheta}^\text{T}\,\mathbf{m_\text{ord}}, & \text{if }  \tilde{d}_{n} \text{ is unreliable},
\end{array}\right.
\end{equation}

\noindent where $\bm{\vartheta}$ and $\mathbf{m_\text{ord}}$ vectors are scaled as $i \times 1$ and $\vartheta_i^0$ is the total number of zeros in the vector $\bm{\vartheta}$ until the $i$th step. Therefore, for $K=5$ and $\bm{\vartheta}^\text{T} = \left[1\,0\,1\,1\,0\right]$, we have

\begin{equation}
\left\{ \begin{array}{ll}
    \text{for } i=1,& d_\text{VGL} = M - K + \bm{\vartheta}^\text{T}\,\mathbf{m_\text{ord}} \\
\text{for } i=2,& d_\text{VGL} = M - K + \bm{\vartheta}^\text{T}\,\mathbf{m_\text{ord}} + 1 \\
\text{for } i=3,& d_\text{VGL} = M - K + \bm{\vartheta}^\text{T}\,\mathbf{m_\text{ord}} + 1\\
\text{for } i=4,& d_\text{VGL} = M - K + \bm{\vartheta}^\text{T}\,\mathbf{m_\text{ord}} + 1 \\
\text{for } i=5=K,& d_\text{VGL} = M - K + \bm{\vartheta}^\text{T}\,\mathbf{m_\text{ord}} + 2.
\end{array}\right.
\end{equation}

Thus, considering the internal list, the diversity order achieved by AA-VGL-DF is $M-K +\left(\bm{\vartheta}^\text{T}\,\mathbf{m_\text{ord}} + \vartheta^0\right)$,

The external list also contributes to the diversity gain. As the external list is comparable as a low complexity ML detector, the increase gain in the diversity order can follow the same idea. As the ML detector has a diversity gain of $M$\cite{LZheng2003} and the size of the group list is variable, we consider the $\nu$ number of symbols chosen to be verified in a total of $\mathcal{G}$ possible vectors. In this way, the diversity order achieved by the AA-VGL-DF is given by~(\ref{eq:div_ord}).

%%%%%%%%%%%%%%%%%%%%%%%%%%%%%%%%%%%%%%%%%%%%%%%%%%%%%%%%%%%%%%%%%%%%%%%%%%%%%%%%%
% SIMULATIONS
%%%%%%%%%%%%%%%%%%%%%%%%%%%%%%%%%%%%%%%%%%%%%%%%%%%%%%%%%%%%%%%%%%%%%%%%%%%%%%%%%
\section{Numerical Results} \label{sec:sim}

In this section, we evaluate the performance of the AA-VGL-DF and other relevant mMTC detection schemes. We consider an underdetermined mMTC system with $N=128$ devices and a single base-station equipped with $M=64$ antennas. The evaluated schemes experience an independent and identically-distributed (i.i.d.) random flat-fading channel model and the values $a_{m,n}$ of (\ref{eq:channel}) are taken from complex Gaussian distribution of $\mathcal{C}\mathcal{N}\left(0,1\right)$. The active devices radiate QPSK symbols with power values drawn uniformly at random in $\left[0.1,0.3\right]$ and the activity probabilities are given by a beta-binomial distribution, as described in Section~\ref{sec:syst_model}. Each transmission slot has 128 symbols, split into 60 metadata and 68 data. This balance between pilots and data is suggested in~\cite{LLiu2018}. For systems that need explicit channel estimation, we considered the scheme described in Section~\ref{subsec:ul}.

\begin{figure}[t]
    \begin{center}
        \includegraphics[scale=0.9]{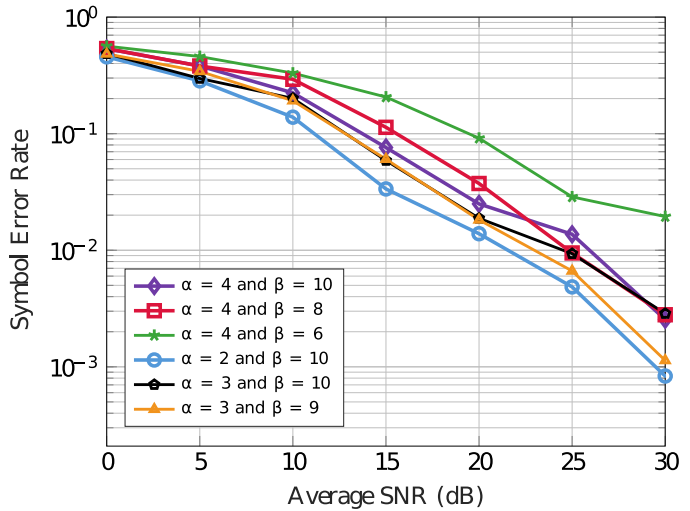}
        \caption{\footnotesize Symbol Error Rate vs. Average SNR of the AA-VGL-DF algorithm in different sparsity scenarios. The activity pattern of devices is determined by a random variable with beta distribution, as shown in Fig.~\ref{fig:prob_act_devices}. In the legend, are shown the $\alpha$ and $\beta$ parameters of each considered distribution, for $N=128$ and $M=64$.}
        \label{fig:NSER_diff_esp}
    \end{center}
\end{figure}

\begin{figure}[t]
    \begin{center}
        \includegraphics[scale=0.9]{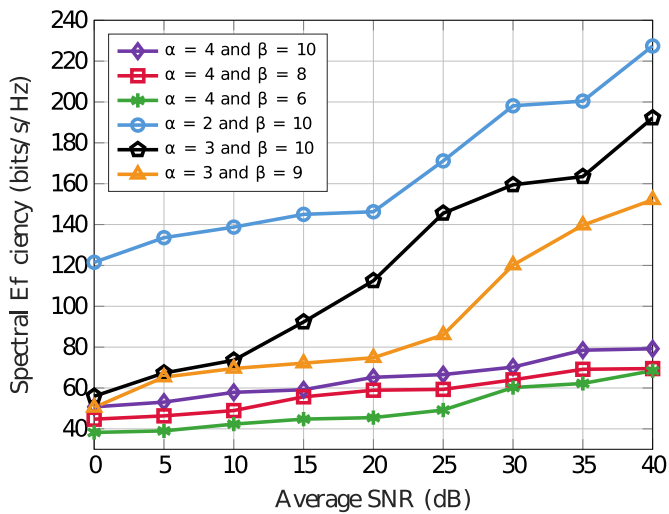}
        \caption{\footnotesize Spectral Efficiency vs. Average SNR of the AA-VGL-DF algorithm with imperfect CSI in different sparsity scenarios. In the legend, are shown the $\alpha$ and $\beta$ parameters of each considered distribution, for $N=128$ and $M=64$.}
        \label{fig:result_Sp_Eff_VGL}
    \end{center}
\end{figure}

Initially, we verify the Symbol Error Rate (SER) performance and the Spectral Efficiency of the AA-VGL-DF for the six sparsity scenarios shown in Fig.~\ref{fig:prob_act_devices}. Fig.~\ref{fig:NSER_diff_esp} shows that as lower is the activity probability of devices, better is the SER performance of AA-VGL-DF.  The result of Fig.~\ref{fig:result_Sp_Eff_VGL} illustrates the achievable spectral efficiency of the system with the AA-VGL-DF detection scheme and shows that, as the sparsity increases, the spectral efficiency also increases. This is due to the reduced number of block collisions and better detection performance, thus reducing the interference. Both plots consider the average SNR as $10\log\left(N\, \sigma^2_x/\sigma^2_v\right)$.

Given the SER results of those different scenarios, we choose the beta-distribution with $\alpha = 4$ and $\beta=8$ as it provides an intermediary sparsity, to compare the SER and Bit Error Rate (BER) performances of the AA-VGL-DF and other relevant mMTC detection schemes.

The numerical results of both uncoded and coded systems are averaged over $10^5$ runs. The performance of AA-VGL-DF is compared with other relevant schemes, as the linear mean squared error (LMMSE), unsorted SA-SIC~\cite{Knoop2014}, SA-SIC with A-SQRD~\cite{Ahn2018}, AA-RLS Linear, AA-RLS-DF~\cite{DiRennaWCL2019} and a version of AA-RLS-DF with the internal list of this work. Besides that, we analyze a version with AA-VGL-DF with perfect activity user detection (AUD) and, as a lower bound, the Oracle LMMSE detector, which has the knowledge of the index of nonzero entries, is considered.

\begin{figure}[t]
    \begin{center}
        \includegraphics[scale=0.9]{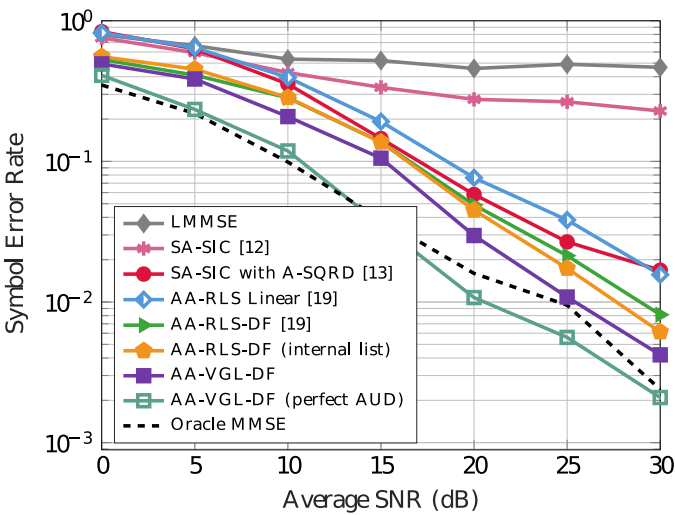}
        \caption{\footnotesize Symbol Error Rate values vs. Average SNR. Parameters of proposals are $\lambda = 0.92$, $\gamma = 0.001$ and $\xi=10$. The pattern activity of the $N=128$ devices is modelled with a beta-binomial distribution with $\alpha=4$ and $\beta=8$. We consider imperfect CSI in the approaches which depends of the channel estimation.}
        \label{fig:result_SER}
    \end{center}
\end{figure}

\begin{figure}[t]
    \begin{center}
        \includegraphics[scale=0.9]{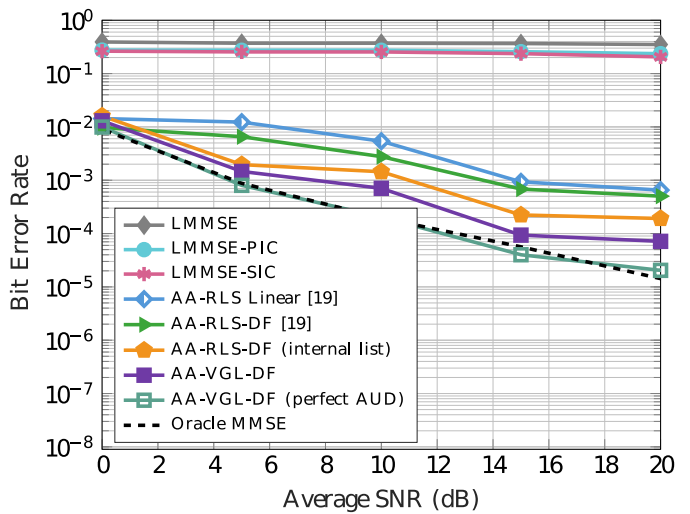}
        \caption{\footnotesize Bit Error Rate values vs. Average SNR. LDPC with block length of 128, symbol rate $R = 0.5$, refined by 2 decoding iterations for the same scenario of Fig.~\ref{fig:result_SER}.}
        \label{fig:result_BER}
    \end{center}
\end{figure}

\begin{figure}[t]
    \begin{center}
        \includegraphics[scale=0.9]{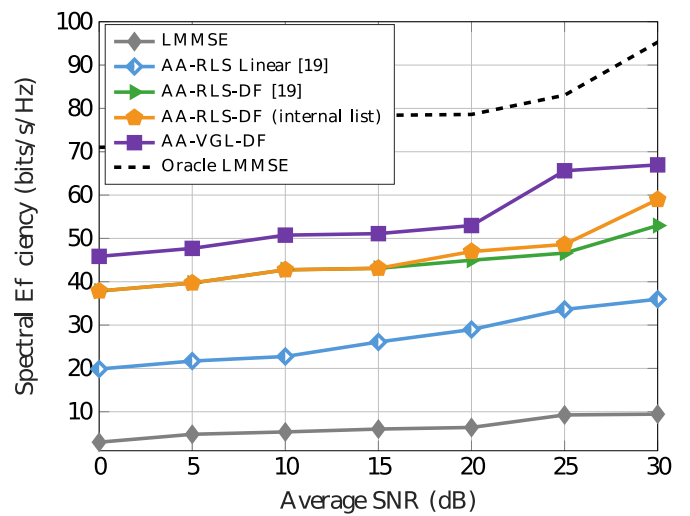}
        \caption{\footnotesize Spectral Efficiency vs. Average SNR. Parameters of proposals are $\lambda = 0.92$, $\gamma = 0.001$ and $\xi=10$. The pattern activity of the $N=128$ devices is modelled with a beta-binomial distribution with $\alpha=4$ and $\beta=8$ and imperfect CSI.}
        \label{fig:result_SR}
    \end{center}
\end{figure}

Fig.~\ref{fig:result_SER} shows the symbol error rate performance of the considered algorithms. LMMSE has a poor performance as the system is underdetermined. Due to error propagation, the unsorted SA-SIC does not perform well. SA-SIC with A-SQRD is effective since it considers the activity probabilities, but under imperfect CSI conditions, its performance is not so good. In contrast, as AA-RLS-DF does not need explicit channel estimation, it is more efficient. The decision-feedback scheme provides a SER gain due to the interference cancellation, which also happens by including the internal list. The proposed schemes with lists of candidates obtain results that outperform the other relevant schemes, approaching the lower bound. The AA-VGL-DF with perfect AUD surpasses the lower bound for high SNRs, where the filter weights are better adjusted and the list schemes are able to correct more errors.

For the coded systems with IDD, Fig.~\ref{fig:result_BER} shows the BER of the already considered algorithms under the scheme proposed in Section~\ref{sec:prop_it_det_dec}. The LDPC matrix has 256 columns and 128 rows, avoiding length-4 cycles and with 6 ones per column. The Sum-Product Algorithm (SPA) decoder is used and the average SNR is $10\log\left(N R\, \sigma^2_x/\sigma^2_v\right)$, where $R=1/2$ is the rate of the LDPC code. The sparsity of the mMTC approach degrades the expected efficiency of LMMSE-PIC, obtaining little variation in relation to LMMSE and LMMSE-SIC. The hierarchy of performance of the other considered algorithms is the same as the uncoded case but with better error rate values. The iterative scheme matches the results for low bit error rate values. Fig.~\ref{fig:result_SR} exhibits the spectral efficiency of the considered algorithms. The filter refinement promoted by the internal and external lists provokes a better spectral efficiency than the other detection schemes. The oracle LMMSE is the upper bound of the system.

%%%%%%%%%%%%%%%%%%%%%%%%%%%%%%%%%%%%%%%%%%%%%%%%%%%%%%%%%%%%%%%%%%%%%%%%%%%%%%%%%
% CONCLUSION
%%%%%%%%%%%%%%%%%%%%%%%%%%%%%%%%%%%%%%%%%%%%%%%%%%%%%%%%%%%%%%%%%%%%%%%%%%%%%%%%%
\section{Concluding remarks} \label{sec:Conc}

In this paper, we have proposed and investigated the AA-VGL-DF detection scheme, for mMTC. Considering different sparsity scenarios, we have presented a list-based DF detector along with an $l_0$-norm regularized RLS algorithm. In order to mitigate error propagation, we employ two lists schemes, based on constellation points that generate candidates for detection. Simulations have shown that AA-VGL-DF significantly outperforms existing approaches with a competitive computational complexity. AA-VGL-DF is also compared and analysed in terms of spectral efficiency and diversity. We have also incorporated into AA-VGL-DF an IDD scheme based on LDPC codes modified to the mMTC scenario.

\end{document}